\documentstyle[12pt]{article}
\renewcommand{\thefootnote}{\fnsymbol{footnote}}
\begin{document}
\begin{flushright}
Columbia preprint CU--TP--756
\end{flushright}
\vspace*{1cm}
\setcounter{footnote}{1}
\begin{center}
{\Large\bf The Time--Delay Signature of Quark--Gluon--Plasma
Formation in Relativistic Nuclear Collisions \footnote{ This 
work was supported by the Director, Office of Energy
Research, Division of Nuclear Physics of the Office of 
High Energy and Nuclear Physics of the U.S.\ Department of 
Energy under Contract No.\ DE-FG-02-93ER-40764.}}
\\[1cm]
Dirk H.\ Rischke and Miklos Gyulassy \\ ~~ \\
{\small Physics Department, Pupin
Physics Laboratories, Columbia University} \\ 
{\small 538 W 120th Street, New
York, NY 10027, U.S.A.} \\ ~~ \\ ~~ \\
{\large June 1996}
\\[1cm]
\end{center}
\begin{abstract} 
The hydrodynamic expansion of quark--gluon plasmas
with spherical and longitudinally boost-invariant geometries 
is studied as a function of the initial energy density.
The sensitivity of the collective flow pattern to uncertainties in 
the nuclear matter equation of state is explored. 
We concentrate on the effect of a possible finite width, 
$\Delta T \sim 0.1\, T_c$, of the transition region between 
quark--gluon plasma and hadronic phase.
Although slow deflagration solutions that act 
to stall the expansion do not exist for $\Delta T > 0.08\, T_c$,
we find, nevertheless, that the equation of state
remains sufficiently soft in the transition region to delay the
propagation of ordinary rarefaction waves for a considerable time.
We compute the dependence of the pion--interferometry correlation
function on $\Delta T$, since this is the most promising 
observable for time-delayed expansion. The signature of 
time delay, proposed by Pratt and Bertsch, is an enhancement
of the ratio of the inverse width of the pion correlation function 
in out--direction to that in side--direction. 
One of our main results is that this generic signature of 
quark--gluon plasma formation is rather robust
to the uncertainties in the width of the transition region.
Furthermore, for longitudinal boost-invariant geometries,
the signal is likely to be maximized around RHIC energies,
$\sqrt{s} \sim 200$ AGeV. 
\end{abstract}
\newpage
\renewcommand{\thefootnote}{\arabic{footnote}}
\setcounter{footnote}{0}
\section{Introduction}

Lattice calculations of the thermodynamical functions of
quantum chromodynamics (QCD) indicate \cite{QM96} 
that, (at zero net baryon density) in the vicinity of a 
critical temperature $T_c \sim 160$ MeV, strongly interacting
matter undergoes a rapid transition from a (chirally broken, confined) 
hadronic phase to a (chirally symmetric, deconfined) quark--gluon plasma
(QGP). The width of that transition region is presently only known 
to be in the range $0 \leq \Delta T < 0.1 \, T_c \sim 16$ MeV.
Therefore, one cannot yet conclude whether the transition is a first 
order phase transition ($\Delta T=0$), or merely a rapid increase 
of the entropy density associated with the change from $d_H$ hadronic to
$d_Q$ quark and gluon degrees of freedom. 

One of the primary goals of relativistic heavy-ion physics 
is the creation and experimental observation of the predicted
QGP phase of matter. Many signatures have been proposed
such as electromagnetic radiation of thermal dileptons and 
photons \cite{dilepton}, $J/\Psi$--suppression \cite{JPsi}, 
jet quenching \cite{jetquench}, strangelet formation \cite{strangelet}, or
disordered chiral condensates (DCC's) \cite{wilcek}. 
These signatures, however, do not depend directly on the actual 
form of the nuclear matter equation of state.
Thermal electromagnetic radiation is, for instance, generic to any
hot system, independent from its degrees of freedom (as long
as they have electromagnetic charge).
For example, it was shown \cite{kapusta} that a hot hadron
gas shines as brightly as a QGP. Similarly, jet quenching and
$J/\Psi$--suppression\footnote{Renewed interest in that signature,
has, however, emerged with the advent of recent results from
the NA50 experiment at the CERN SPS, showing a rapid variation 
of $J/\Psi$--suppression as a function of transverse energy in
Pb+Pb--collisions at 158 AGeV  \cite{QM96NA50}.} 
are generic consequences of final state interactions
in any form of dense matter \cite{huefner}. 
Finally, strangelet or DCC formation require very specific
assumptions about the dynamical evolution of the system.

It is therefore of interest to study signals that are more
directly related to the QCD equation of state. 
Signals of this type emerge from the influence of the equation 
of state on the {\em collective\/} dynamical evolution of 
the system. Such signals must be calculated
within the framework of relativistic hydrodynamics \cite{strottman}, 
since that is the only dynamical model which provides a {\em direct\/} 
link between collective observables and the equation of
state. Of course, the use of such an approach requires a strong dynamical
assumption, namely that the equilibration rates are much larger
than typical gradients of thermodynamic quantities in the system. 
It now appears that radiative gluon energy loss in a QGP may
be sufficiently large \cite{eloss} to support local equilibration
on time scales less than 1 fm/c. In the following, we
therefore neglect dissipative effects and assume the validity of 
ideal hydrodynamics to compute the collective evolution of the system. 

It was shown in \cite{shuryak,test1,dhrmg} 
that the transition to the QGP softens the equation of
state in the transition region, and thus reduces the tendency
of matter to expand on account of its internal pressure.
This, in turn, delays the expansion and considerably
prolongs the lifetime of the system.
In \cite{csernai,puersuen} it was moreover shown that
this effect leads to a reduction of the transverse directed
flow in semi-peripheral collisions that can be readily tested
experimentally at fixed target energies \cite{E877}. 

In \cite{dhrmg} the flow structure for the one--dimensional
expansion of a slab of matter (``Landau expansion model''
\cite{landau}) was studied as a function of its
initial energy density $\epsilon_0$ and of the width $\Delta T$ of
the transition region in the equation of state.
In particular, it was shown how, in addition to 
the softening of the equation of state,
for $\Delta T < \Delta T^* \simeq 0.07676\, T_c$ the type of the
hydrodynamical expansion solution changes qualitatively from
a simple rarefaction wave to a rarefaction discontinuity.
Such deflagration fronts \cite{vanH} have small velocities,
$v_{\rm d} \ll c$, and thus the conversion of QGP matter
into hadronic matter is considerably delayed.
In \cite{dhrmg} we also showed that for larger transition widths
($\Delta T > \Delta T^*$) the lifetime of matter with temperature
around $T_c$ is considerably reduced but cooler regions with 
$T \sim 0.7\, T_c$ remain long-lived.

In the present work we extend the investigations of 
\cite{dhrmg} to more realistic, 3--di\-men\-sio\-nal
geometries. We consider systems with spherical
symmetry (``fireball'' geometry), and with transverse cylindrical symmetry 
and boost-invariant initial conditions in longitudinal direction
(``Bjorken cylinder'' geometry \cite{bjorken}). 
For these symmetric geometries it is possible
to use a modified one--dimensional algorithm as e.g.\ 
presented in \cite{test1}.
Note that in contrast to the simple one--dimensional Landau geometry, 
single inclusive particle spectra and also two--particle correlation functions
can be calculated in a realistic way from the hydrodynamic solutions 
for the above geometries.

In this paper we show that the influence of the width $\Delta T$ of the 
transition on the dynamical evolution of the system in these 3--dimensional
geometries is very similar to that found in \cite{dhrmg}
for purely one--dimensional expansion. In particular,
the softening of the equation of state
associated with a rapid cross-over region
leads to a delay in the expansion and a prolonged lifetime
of matter with temperature {\em below\/} $T_c$.  

To explore observable consequences of the prolonged lifetime of the
system, we then calculate the inverse width of the
two--particle correlation functions in side-- and out--direction,
$R_{\rm \,side}$ and $R_{\rm \,out}$. We demonstrate that
the initial energy density dependence of the system's 
lifetime is mapped closely by the excitation function of the ratio 
$R_{\rm \, out}/R_{\rm \, side}$. The phase transition
leads to a time-delayed expansion that is reflected in an enhancement of 
the above ratio relative to its value obtained in the
ideal gas case without a transition. This signature of the QGP transition
was first proposed by Pratt \cite{pratt}
who performed similar calculations for the fireball geometry\footnote{Pratt's
method to solve the hydrodynamical equations and to calculate the 
correlation functions differ in important aspects from our approach, 
but our final results for the correlation functions turn out to be 
rather similar to his.}
and, independently, by Bertsch et al.\ \cite{bertsch} 
who employed a kinetic model for the hadronization
in the Bjorken cylinder geometry.
These previous investigations assumed the existence of a sharp
first order transition, $\Delta T=0$.
We show here that this signature
is a generic feature of a rapid cross-over transition
and holds for $\Delta T=0.1\, T_c$ as well.

The importance of the time delay as a signature of QGP formation
is that it is among the few signatures that probe directly the
equation of state of strongly interacting matter. It tests 
whether there exists a ``soft region'' in the equation of state
of ultra-dense matter \cite{shuryak} which 
is most naturally characterized \cite{dhrmg} by a reduction of the local
velocity of sound $c_S$ as a function of the local energy density.
As discussed in detail in the next section,
a very general feature of a transition (that happens within a sufficiently
narrow temperature interval $\Delta T$) is the existence of such a region. 
For matter passing through that region of energy densities during the
expansion phase, the flow will temporarily slow down
or even possibly stall under suitable conditions. 
The main result of the present work is
to identify initial conditions where the resulting stall
is sufficiently long to be observable via pion interferometry
\cite{pratt,bertsch}. As we show, the maximum time delay
may occur at AGS as well as at RHIC energies, depending on whether matter
is initially stopped in a fireball or expands as in Bjorken's 
longitudinally boost-invariant scenario.

The remainder of this paper is organized as follows.
In Section 2 we present the equation of state used in
our investigations and explain our method to solve 
the hydrodynamical equations numerically for the
above mentioned geometries. In Section 3 we compare the
hydrodynamical expansion solutions for different initial
energy densities, for different values of $\Delta T$, and for
different values of degrees of freedom in the QGP and hadronic phase.
Section 4 is devoted to the discussion of the lifetime
of the system and how it can be measured via the widths of the side-- 
and out--correlation function.
Section 5 concludes with a summary and discussion 
of our results. Two Appendices contain the derivation
of the formulae employed to calculate the 
two--particle correlation functions in the hydrodynamical
framework. Natural units $\hbar = c = k_B = 1$ are used throughout this
paper.

\section{Equation of State and Hydrodynamics}

Available lattice data for the entropy density in full QCD
can be approximated by the simple parametrization
\cite{dhrmg,blaizot}
\begin{equation} \label{eos}
\frac{s}{s_c}(T) = \left[\frac{T}{T_c}\right]^3 
\left( 1 + \frac{d_Q-d_H}{d_Q+d_H} \, 
\tanh \left[ \frac{T-T_c}{\Delta T} \right] \right)~,
\end{equation}
where $s_c = const. \times \frac{1}{2}\, (d_Q+d_H)\, T_c^3$ 
is the entropy density at $T_c$.
Pressure $p$ and energy density $\epsilon$ follow
then from  thermodynamical relationships.
For $\Delta T=0$, the equation of state (\ref{eos}) reduces
to the MIT bag equation of state \cite{MIT} with bag constant
$B= \frac{1}{2}\, (d_Q/d_H-1)\, T_c\, s_c /\,(d_Q/d_H+1)$.
If one measures energies in units of $T_c$ and
energy densities in units of the
enthalpy density $\epsilon_c +p_c =
T_c\, s_c$, the equation of state (\ref{eos})
depends only on the ratio $d_Q/d_H$, and not on $d_Q$ and $d_H$
separately. For $\Delta T =0$, this ratio determines the
latent heat (density) $\epsilon_Q - \epsilon_H \equiv 4 B$.
(Here, $\epsilon_Q = \frac{1}{2}\, (4\, d_Q/d_H -1)\, T_c\, s_c\,
/\,(d_Q/d_H+1)$ is the energy density at the phase boundary between
mixed phase and QGP, $\epsilon_H = \frac{3}{2}\, T_c\, s_c\,
/\,(d_Q/d_H+1)$ is that at the boundary between mixed and hadronic phase.)

For the case $d_H =3$ (corresponding to an ultrarelativistic gas 
of pions) and $d_Q = 37$ (corresponding to two massless flavours
of quarks and antiquarks, and eight massless gluons), 
the latent heat, $4B=1.7\, T_c\, s_c \simeq 1.272 $ GeV fm$^{-3}$, is large.
On the other hand, including a resonance gas 
in the hadronic phase and/or reducing
the effective number of degrees of freedom
on the QGP side \cite{neumann}, $d_Q/d_H = 3$ may be 
taken as a (perhaps more realistic) lower limit, with a smaller latent heat 
$\epsilon_Q - \epsilon_H = T_c\, s_c$.
Assuming that the high-temperature phase consists of gluons
only (such as expected for the ``hot-glue scenario''
\cite{glue}) this would then
correspond to about $400$ MeV fm$^{-3}$ in physical units.
To cover the range of uncertainty in the QCD equation of state
we consider $d_Q/d_H = 37/3$ and $d_Q/d_H = 3$, and 
$\Delta T = 0$ and $0.1\, T_c$ in our investigation
of hydrodynamical expansion solutions.

In Fig.\ 1 we show (a) the entropy density and (b) the energy density 
as functions of temperature, and (c) the pressure 
and (d) the velocity of sound squared $c_S^2 \equiv {\rm d}p/
{\rm d}\epsilon$ as functions of energy density
for $\Delta T=0$, $0.1\, T_c$, and an ideal gas with $d_H$ degrees
of freedom for $d_Q/d_H = 37/3$. The corresponding plot for $d_Q/d_H = 3$ 
looks rather similar qualitatively and is not shown.
Quantitative differences are: (i) the phase
boundaries are different, $\epsilon_Q$ shifts from 
$1.8125\,T_c\,s_c$ to $1.375\, T_c\,s_c$, $\epsilon_H$ from
$0.1125\, T_c\,s_c$ to $0.375\, T_c\, s_c$, and
$p_c$ from $0.0375\, T_c\, s_c$ to $0.125\, T_c\, s_c$. Also,
as mentioned above, (ii) the latent heat $\epsilon_Q-
\epsilon_H$ is reduced.

Figs.\ 1 (a,b) present the thermodynamic functions in
a form to facilitate comparison with lattice data. 
Present lattice data for full QCD can be
approximated with a choice of $\Delta T$ in 
the range $0 \leq \Delta T < 0.1\, T_c$. In the
hydrodynamical context, however, Figs.\ 1 (c,d) are more relevant. 
As can be seen in (c), for $\Delta T=0$ the pressure stays constant,
$p_c = \frac{1}{2}\, T_c\, s_c\, /\, (d_Q/d_H+1)$, in the
mixed phase $\epsilon_H \leq \epsilon \leq \epsilon_Q$.
Hydrodynamical expansion is, however, driven by 
pressure {\em gradients}. It is therefore the velocity 
of sound, Fig.\ 1 (d), that is the most relevant measure of 
the system's tendency to expand. It represents the capability
to perform mechanical work (which is proportional to pressure
gradients ${\rm d}p$) for a given gradient in energy
density ${\rm d} \epsilon$. For $\Delta T =0$,
the velocity of sound vanishes in the mixed phase, i.e.,
mixed phase matter does not expand at all on its own account, 
even if there are strong gradients in the energy density.
This has the consequence that it does not perform 
mechanical work and therefore cools less rapidly.
For finite $\Delta T$, pressure gradients are finite, but still
smaller than for an ideal gas equation of state, and
therefore the system's tendency to expand 
is also reduced, cf.\ Fig.\ 1 (d). 

Hydrodynamics is defined by local energy--momentum conservation,
\begin{equation} \label{eom}
\partial_{\mu} T^{\mu \nu} = 0~.
\end{equation}
Under the assumption of local thermodynamical equilibrium (the so-called
``ideal fluid'' approximation) the energy--momentum tensor
$T^{\mu \nu}$ assumes the particularly simple form \cite{LL}
\begin{equation} \label{tmunu}
T^{\mu \nu} = (\epsilon + p)\, u^{\mu} u^{\nu} - p\, 
g^{\mu \nu}~,
\end{equation}
where $u^{\mu} = \gamma\, (1,{\bf v})$ is the 4--velocity
of the fluid (${\bf v}$ is the 3--velocity, 
$\gamma \equiv (1-{\bf v}^2)^{-1/2}$,
$u_{\mu} u^{\mu} = 1$), and $g^{\mu \nu} = {\rm diag}(+,-,-,-)$
is the metric tensor. The system of equations (\ref{eom}) is closed 
by choosing an equation of state
in the form $p=p(\epsilon)$, i.e., as depicted in Fig.\ 1 (c).
In the ideal fluid approximation, the (equilibrium) equation of 
state is the {\em only\/} input to the hydrodynamical 
equations of motion (\ref{eom}) that relates to properties of
the matter under consideration and is thus able to influence
the dynamical evolution of the system. The final results are
uniquely determined once a particular initial condition and a 
decoupling (``freeze-out'') hypersurface are specified.

The symmetry of the fireball and Bjorken cylinder geometry
affects that the system of four equations (\ref{eom}) 
reduces to two independent equations. With the definition
$E \equiv T^{00}, \, M \equiv T^{0r}$, where the index $r$
indicates the radial component of the energy--momentum tensor,
the respective equations read
\begin{eqnarray} \label{eom2a}
\partial_t\, E + \partial_r\, [(E+p)v] & = & 
-\, F(E,p,v,r,t) \, \, , \\
\partial_t\, M + \partial_r\, (Mv+p) & = & 
-\, G(M,v,r,t) \, \, . \label{eom2b}
\end{eqnarray}
Here, $v$ is the radial component of the velocity. For
the fireball geometry, $F$ and $G$ do not depend on the time $t$
explicitly,
\begin{equation} \label{FGfb}
F_{\rm fb} (E,p,v,r) = \frac{2\, v}{r}\, (E+p)\,\, , 
\,\,\,\, 
G_{\rm fb} (M,v,r) = \frac{2\, v}{r}\, M\,\, .
\end{equation}
For the Bjorken cylinder geometry, the above equations describe the
system's transverse evolution at $z=0$ 
(cf.\ \cite{baym}), and due to the assumption of
longitudinal boost invariance \cite{bjorken}, 
the hydrodynamical solution for
arbitrary $z$ can be easily obtained by a Lorentz boost
with (space--time) rapidity $\eta = {\rm Artanh}\, [z/t]$.
The functions $F$ and $G$ read in this case
\begin{equation}
F_{\rm Bj} (E,p,v,r,t) = \left( \frac{v}{r} + \frac{1}{t}
\right) (E +p)\,\, , \,\,\,\,
G_{\rm Bj}(M,v,r,t) = \left( \frac{v}{r} + \frac{1}{t} \right) M\,\, .
\end{equation}
In order to solve (\ref{eom2a},\ref{eom2b}), we employ Sod's
operator splitting method \cite{sod}, i.e., for each time step we
first generate solutions of the hydrodynamical equations for $F=G=0$.
In this form, the equations are purely one--dimensional and
can therefore be solved with e.g.\ the relativistic 
HLLE scheme presented in \cite{test1,schneider}. 
The performance of this algorithm for solving one--dimensional 
hydrodynamical problems and with equations of state 
featuring phase transitions was shown to be excellent
\cite{test1,test2}. It has also been employed in
\cite{dhrmg} to solve the Landau expansion problem.

In a second step, Sod's method prescribes to correct
$E$ and $M$ for the respective
geometry by solving the ordinary differential equations
\begin{equation}
\frac{{\rm d}E}{{\rm d}t} = - F(E,p,v,r,t)\,\, , \,\,\,\,
\frac{{\rm d}M}{{\rm d}t} = - G(M,v,r,t)\,\, .
\end{equation}
More specifically, in a slight variation of
Sod's method we solve the finite difference equations
\begin{equation} \label{correct}
E = \tilde{E} - F(\tilde{E},\tilde{p},\tilde{v},r,t) 
\, {\rm d}t\,\, , \,\,\,\,
M =  \tilde{M} - G(\tilde{M},\tilde{v},r,t) 
\, {\rm d}t\,\, ,
\end{equation}
where quantities with a tilde are the solutions of
the hydrodynamical equations with $F=G=0$ generated 
previously with the relativistic HLLE scheme. This two--step
predictor--corrector scheme is repeated for each time step to
obtain the complete time evolution of the system.

Fig.\ 2 shows temperature and laboratory energy density
profiles calculated with Sod's method for the expansion of 
(a,b) a sphere and (c,d) a Bjorken cylinder 
(for initial time $t_0 \equiv \tau_0 = 0.1\, R$) with an ideal
gas equation of state $p=\epsilon/3$ in comparison
to profiles generated with the semi-analytic method
of characteristics \cite{baym}. That method is a benchmark
test for numerical algorithms as long as the hydrodynamical
solution is continuous \cite{test1}.
The grid spacing for the HLLE scheme is taken as 
$\Delta x = 0.01\, R$, the time step width for the
HLLE scheme and the corrector step (\ref{correct}) is
$\Delta t = 0.99\, \Delta x$.
As one observes, agreement is excellent (even on a
logarithmic scale) which gives
confidence that Sod's method as described above
works for the more complicated equation of state 
(\ref{eos}) as well
(which leads to discontinuous hydrodynamical expansion solutions
for $\Delta T < \Delta T^* \simeq 0.07676\, T_c$ \cite{test1}).

\section{Expansion solutions}

In this section we present the hydrodynamical
expansion solutions for the fireball and Bjorken cylinder
geometry for different initial energy densities
$\epsilon_0$. We assume $\epsilon_0$ to be homogeneous
throughout the system. We compare solutions for the equation
of state (\ref{eos}) with $\Delta T=0$ and $0.1\, T_c$ 
with solutions for an ideal gas of equation of state
with $d_H$ degrees of freedom. We explicitly show results
for $d_Q/d_H=37/3$ and, where necessary, 
comment on the difference to the case $d_Q/d_H=3$.
As in \cite{dhrmg} we will use the notion ``lifetime''
for the intercept of a particular isotherm with
the $t$--axis. 

Note that in our comparison we fix the initial {\em energy density\/}
rather than the initial temperature or entropy density, as sometimes 
assumed \cite{kataja}, because the latter are derived
thermodynamic quantities while the former is determined by
non-equilibrium energy loss mechanisms. For instance, at RHIC energies
it is expected \cite{glue} that semi-hard perturbative QCD processes
determine the initial energy density in the range
$\epsilon_0 \sim 10-20$ GeV/fm$^3$.

\subsection{Fireball geometry}

This case is rather similar to the one--dimensional Landau 
expansion studied in \cite{dhrmg}. Differences are
solely due to the spherical geometry (i.e., due to
the extra terms (\ref{FGfb}) in the equations of motion).

In Fig.\ 3 we show temperature profiles and isotherms
in the $t-r$ plane for $\epsilon_0 = \epsilon_H = 
0.1125\, T_c\, s_c$. Of course, for this initial
energy density, the case $\Delta T=0$ (a,b) is identical to
the ideal gas case (e,f). Note the delayed expansion in
the case $\Delta T=0.1\, T_c$, Figs.\ 3 (c,d), due to the reduction in
the velocity of sound, cf.\ Fig.\ 1 (d). This is quite
similar to the one--dimensional expansion (cf.\ Figs.\ 3 (c,i) 
of Ref.\ \cite{dhrmg}), although the overall scale of the lifetimes
is now considerably reduced since the system has two more spatial 
dimensions into which it can expand.

Fig.\ 4 shows the situation for an initial energy density
$\epsilon_0 = 1.875\, T_c \, s_c$ which is close to 
$\epsilon_Q = 1.8125\, T_c\, s_c$. As in the one--dimensional
expansion (cf.\ Fig.\ 5 of \cite{dhrmg}), 
the lifetime of the system is very long in the case
$\Delta T=0$, due to the small velocity of the deflagration
front converting mixed phase matter into hadrons. 
As compared to the lifetime of
the mixed phase in the one--dimensional expansion, $t_{\rm \,life}
\sim 40\, R$ (cf.\ Fig.\ 5 (g) of \cite{dhrmg}), 
that lifetime is now, however, only about $18\, R$,
due to the more rapid expansion in three spatial dimensions. 
As in the one--dimensional
case, cooling is faster for $\Delta T=0.1\, T_c$ (cf.\ Fig.\ 
5 (i) of \cite{dhrmg}). While in that case, however, the lifetimes were
only slightly shorter than for $\Delta T=0$, now they are
reduced by about a factor of 3.
The fastest expansion is that for the ideal gas,
in accord with the discussion of Fig.\ 1 (d).

In Fig.\ 5 the initial energy density is well above
the mixed phase region, $\epsilon_0 = 18.75\, T_c\, s_c \sim 10\,
\epsilon_Q$. As for the analogous one--dimensional 
expansion solution (Fig.\ 7 of \cite{dhrmg})
the internal energy density and pressure are so large that, in the
case of a transition to the QGP, Figs.\ 5 (a--d),
the system explodes rather than 
burns slowly as in the preceding case, Fig.\ 4.
Nevertheless, the expansion is still somewhat delayed 
as compared to the expansion of an ideal gas, Figs.\ 5 (e,f).

\subsection{Bjorken cylinder geometry}

For the discussion of the expansion in the Bjorken cylinder
geometry it is instructive to first focus on purely longitudinal
expansion. Longitudinal boost invariance with regard to
boosts with $\eta = {\rm Artanh}\, [z/t]$ \cite{bjorken}
implies that the longitudinal velocity is given by $v^z = z/t$.
In turn, the longitudinal fluid rapidity is identical with
the boost rapidity $\eta$. It is easy to show that in this case the 
hydrodynamical equations (\ref{eom}) reduce to
\begin{equation} \label{bj}
\left. \partial_{\tau} \, \epsilon \, \right|_{\eta} 
= - \frac{\epsilon+p}{\tau}
\,\, , \,\,\,\, 
\left. \partial_{\eta} \, p \, \right|_{\tau} = 0\,\, ,
\end{equation}
where $\tau \equiv \sqrt{t^2-z^2}$ is the proper time 
associated with fluid elements moving with $v^z = z/t$.
The second equation implies that pressure gradients
vanish along space--time hyperbolas $\tau = const.$
and, for baryon-free matter, that the temperature is
constant on these hyperbolas. The first equation
is an ordinary differential equation on
the space--time hyperbolas and describes the cooling
of the system on account of the longitudinal motion.
For baryon-free matter, the thermodynamical identities 
${\rm d}\epsilon = T\, {\rm d}s$ and $\epsilon + p = T s$
imply the very simple cooling law
\begin{equation}
\left. \partial_{\tau} \, s \, \right|_{\eta} = - \frac{s}{\tau}\,\, ,
\end{equation}
which has the solution $s(\tau) = s_0\, [\tau/\tau_0]^{-1}$,
{\em independent\/} from the underlying equation of state.

For an ultrarelativistic ideal gas equation of state
$p=\epsilon/3$, $\epsilon = const. \times T^4$, the solution 
of the longitudinal boost-invariant expansion problem is
\begin{equation} \label{idgas}
\epsilon(\tau) = \epsilon_0 \, [\tau/\tau_0]^{-4/3} \,\, , \,\,\,\,
T(\tau) = T_0 \, [\tau/\tau_0]^{-1/3} \,\, .
\end{equation}
For the equation of state (\ref{eos}) with $\Delta T =0$,
one can easily solve (\ref{bj}) analytically. If
$\epsilon_0 > \epsilon_Q$ we have
\begin{eqnarray}
\epsilon (\tau) & = & \left\{
           \begin{array}{ll} 
           (\epsilon_0 - B)\, [\tau/\tau_0]^{-4/3} +B
                  \,\, , & \tau_0 \leq \tau \leq \tau_Q\,\, , \\
           (\epsilon_Q + p_c)\, [\tau/\tau_Q]^{-1} - p_c
                  \,\, , &  \tau_Q < \tau \leq \tau_H \,\, ,\\
           \epsilon_H\, [\tau/\tau_H]^{-4/3} 
                   \,\, , & \tau > \tau_H \,\, ,
           \end{array}  \right. \\
T(\tau) & = & \left\{
           \begin{array}{ll} 
           T_0\, [\tau/\tau_0]^{-1/3}
               \,\, , & \tau_0 \leq \tau \leq \tau_Q \,\, ,\\
           T_c  \,\, , &  \tau_Q < \tau \leq \tau_H \,\, ,\\
           T_c\, [\tau/\tau_H]^{-1/3} 
                       \,\, , & \tau > \tau_H \,\, . 
           \end{array}  \right. 
\end{eqnarray}
Here, $\tau_Q= \tau_0\, [(\epsilon_Q-B)/(\epsilon_0-B)]^{-3/4}$
is the (proper) time the system enters the mixed phase,
and $\tau_H = \tau_Q\, d_Q/d_H$ the time corresponding to
entry of the hadronic phase. Note that the time spent in 
the mixed phase is linear proportional to the ratio
$d_Q/d_H$. Also, the system does not cool at all in this phase,
$T=T_c=const.$, due to the fact that no mechanical work is performed (cf.\
the above discussion of Fig.\ 1 (d)). As another consequence,
the energy density does not decrease with $\tau^{-4/3}$
as in the QGP and hadronic phase, but
only proportional to $\tau^{-1}$, precisely as in a 
(one--dimensional) free-streaming expansion (where no 
mechanical work is performed as well).

For $\epsilon_Q \geq \epsilon_0 > \epsilon_H$, the solution
reads
\begin{eqnarray}
\epsilon (\tau) & = & \left\{
           \begin{array}{ll} 
           (\epsilon_0 + p_c)\, [\tau/\tau_0]^{-1} - p_c
                  \,\, , &  \tau_0 \leq \tau \leq \tau_H \,\, ,\\
           \epsilon_H\, [\tau/\tau_H]^{-4/3} 
                   \,\, , & \tau > \tau_H \,\, ,
           \end{array}  \right. \\
T(\tau) & = & \left\{
           \begin{array}{ll} 
           T_c  \,\, , &   \tau_0 \leq \tau \leq \tau_H \,\, ,\\
           T_c\, [\tau/\tau_H]^{-1/3} 
                       \,\, , & \tau > \tau_H \,\, . 
           \end{array}  \right.  \label{T}
\end{eqnarray}
Here, $\tau_H = \tau_0\, (\epsilon_0+p_c)/(\epsilon_H +p_c)$.
For $\epsilon_0 \leq \epsilon_H$, the solution is identical to
(\ref{idgas}). 

For finite $\Delta T$ there is, contrary to expectation, 
also a simple, semi-analytic solution 
which does {\em not\/} require to solve the equations of motion 
(\ref{bj}) explicitly. Due to the fact that the entropy density
behaves as $s(\tau) = s_0\, [\tau/\tau_0]^{-1}$ 
{\rm irrespective\/} of the equation of state, and since
the equation of state (\ref{eos}) establishes (for finite $\Delta T$) a
one-to-one correspondence between temperature and entropy
density, there is also a one-to-one correspondence between
temperature and $\tau$. Once $s(\tau)$ and $T(\tau)$ are known,
one can easily calculate $\epsilon(\tau)$ and $p(\tau)$ from
fundamental thermodynamical relationships.

For illustrative purposes, we show in Fig.\ 6 the time evolution
of (a) energy density, (b) entropy density, (c) pressure, and 
(d) temperature assuming an initial energy density 
$\epsilon_0 = 10\, T_c\, s_c$ and $d_Q/d_H=37/3$. Solid lines are
for $\Delta T=0$, dotted lines for $\Delta T=0.1\, T_c$, and
dashed lines for an ideal gas with $d_H$ degrees of freedom.
The change in the cooling law for the energy density when entering
the mixed phase is clearly visible in Fig.\ 6 (a). Also, 
Fig.\ 6 (b) illustrates that the cooling law for the entropy 
density is independent from the equation of state,
although the magnitude of the entropy density
differs for the different equations of state because of
our assumption of fixed initial energy density.

We mention that for $\Delta T=0.1\, T_c$, the parts
of the system with temperature $T > 0.8\, T_c$
cool {\em faster\/} than for the ideal gas, cf.\ Fig.\ 6 (d).
Only parts cooler than $\sim 0.7\, T_c$ survive longer than
in the ideal gas case. This is in agreement with the
results of \cite{dhrmg} for the Landau expansion
model, where it was concluded that it is therefore less likely that
electromagnetic radiation is a viable signature for the transition,
in contrast to the conclusions of \cite{shuryak}. On the other hand, it was
speculated in \cite{dhrmg} (referring to the conclusions of 
\cite{pratt,bertsch}) that, if the system freezes out at
temperatures $\leq 0.7\, T_c$, the delayed expansion might be observable
in side-- and outwards radii of two--particle correlation functions.
This indeed happens to be the case, as will be shown in the next section.

For given initial energy density $\epsilon_0$ and proper
time $\tau_0$, one can ask
the question, how much longer it takes a system
described by the equation of state (\ref{eos}) to reach a certain
freeze-out temperature $T_f \leq T_c$
in comparison to the expansion of an ideal gas (\ref{idgas}).
Since the ideal gas is supposed to freeze out at the same
temperature $T_f$ and to have $d_H$ massless degrees of freedom,
particle and entropy density at freeze-out are 
the same in both cases. For $\Delta T=0$ one can easily answer
this question analytically:
for $\epsilon_0 \gg B$, the freeze-out time
$\tau_f \equiv \tau(T_f)$ is prolonged by a 
factor $[d_Q/d_H]^{1/4}$. For $d_Q/d_H = 37/3$, 
the delay in cooling is about 87\% (as one can also see in
Fig.\ 6 (d)), for $d_Q/d_H=3$ it is
only about 32\%. Note that this prolongation of
the lifetime of the system is in agreement with
the one for the Landau expansion model at very high 
initial energy densities (cf.\ \cite{dhrmg} and
Fig.\ 10 below). Only for initial
energy densities around $\epsilon_Q$ is the
prolongation of the lifetime larger in the
Landau expansion (see Fig.\ 10) due to the fact that there exists a slow
deflagration solution which delays the expansion even further.
Such a solution does not occur in the purely longitudinal
Bjorken expansion.

This changes when we consider transverse motion as well. In the
following we fix $\tau_0=0.1\, R$. This choice is motivated
by the fact that for Au+Au collisions at RHIC, the transverse
radius of the hot zone is of the order 5 fm, while the
time scale of local equilibration is roughly given by the
energy loss of a parton in strongly interacting matter,
$\tau_{dE/dx} \sim 0.5$ fm \cite{eloss}. We have, however,
also considered the case $\tau_0=0.5\, R$ and $\tau_0 = 1/3\,
T_0$ (which is motivated by an uncertainty principle argument)
and will comment on differences to the choice
$\tau_0 = 0.1\, R$ where necessary.

Let us first consider the case $\epsilon_0 = \epsilon_H =0.1125\, 
T_c\, s_c$, cf.\ Fig.\ 7. Due to the strong initial longitudinal
motion, the system cools rather quickly below temperatures
of $0.5\, T_c$, even before the transverse rarefaction
wave reaches the center of the cylinder. The lifetimes are
therefore solely determined by the longitudinal scaling expansion.
This causes the horizontal parts of the isotherms.
For the $T=0.5\, T_c$ isotherm and $\Delta T=0$ one expects from
eq.\ (\ref{T}) the lifetime
$t_{\rm \,life}= 0.5^{-3} \times \, 0.1 R = 0.8\, R$, in good agreement
with Figs.\ 7 (a,b,e,f). (Due to the numerical dissipation adherent to
any finite difference scheme that solves the hydrodynamical equations,
the system cools slightly faster in the numerical calculation,
cf.\ Figs.\ 7 (b,f).)
The effect of longitudinal cooling is reduced 
for larger values of $\tau_0$, cf.\ (\ref{idgas}). 
For instance, in the case $\tau_0=0.5\, R$ the center still has
$T\simeq 0.6\, T_c$ when the transverse wave reaches it.
The lifetimes for $T > 0.6\, T_c$ are therefore (as expected)
a factor of 5 longer.

In Fig.\ 8 we show the situation for 
$\epsilon_0 = 1.875\, T_c\, s_c \sim \epsilon_Q$. As in the preceding case,
it takes some time before the transverse expansion wave reaches
the center. Thus, the lifetimes of the highest temperatures
in the system are solely determined by the longitudinal
scaling solution presented above (horizontal parts of the
respective isotherms).

It is interesting to note that the lifetimes for
$\Delta T=0$ and this particular $\epsilon_0$ 
are not that exceedingly long as they were
in the case of one--dimensional and spherical expansion. 
These long lifetimes were due to the slow velocity of
the deflagration front. A deflagration solution, however, exists
only for energy densities in the mixed phase, where matter is
thermodynamically anomalous \cite{dhrmg}. Here,
the strong longitudinal motion cools the system quickly
below the respective energy densities, such that the transverse
expansion proceeds as a (comparatively fast) simple rarefaction 
wave instead of the deflagration. Thus, the system cools 
even more quickly and the lifetimes are considerably reduced.

This effect is, however, compensated when higher initial energy densities
are considered.
In Fig.\ 9 we show the solutions for $\epsilon_0 = 18.75\, T_c\, s_c
\sim 10\, \epsilon_Q$. This case is close to
the initial conditions expected from mini-jet production at RHIC energies.
Now the initial energy density is sufficiently high
that for $\Delta T=0$ the plasma enters
the mixed phase at relatively late time ($\tau_Q \sim 3$ fm
for $\tau_0=0.5$ fm, see also Fig.\ 6 (d)).
Therefore, the time spent in the mixed phase 
is long enough in this case to allow a transverse deflagration front
to form, cf.\ Figs.\ 9 (a,b). 
In comparison to the ideal gas case (e,f), where no such solution
exists, the expansion is therefore considerably prolonged. 
Such a time delay is observed also for $\Delta T=0.1\, T_c$, 
Figs.\ 9 (c,d), and is here due to the reduction of the 
velocity of sound in the transition region.

We note that for $\tau_0=0.5\, R$, the initial energy density
where one observes this effect is smaller, $\epsilon_0 \sim 2-3\,
\epsilon_Q$. On the other hand, for $\tau_0 = 1/3\, T_0$
the initial time decreases with increasing initial energy density,
enhancing the effect of longitudinal cooling,
such that $\epsilon_0$ has to be very high ($> 40\, \epsilon_Q$)
for the onset of this effect. Then, however, a steady increase
of the initial energy density affects a steady decrease of
$\tau_0$ such that one observes a prolongation of the lifetimes
over a wider range of energy densities (see also Fig.\ 13 below).
It is furthermore interesting to note that 
this phenomenon occurs at about the same
initial energy densities (in units of $T_c\, s_c$) 
in the case $d_Q/d_H = 3$ (cf.\ Fig.\ 12).

Finally, for very high energy densities, one
recovers the case of the violent explosion, and the lifetimes
are again reduced. This happens at lower initial energy densities
for the case $\tau_0 = 0.5\, R$, since the
longitudinal motion is less effective in cooling the system.
For $\tau_0 = 1/3\, T_0$ we have not found a decrease in
the lifetime up to the highest initial energy
density studied by us, $\epsilon_0 = 300\, T_c\, s_c$.
The reason is, as mentioned above, that the 
increasingly more efficient cooling due to longitudinal motion 
compensates the increasing tendency of the system to 
explode transversally.

\section{Lifetimes and two--particle correlations}

In this section we first discuss the lifetimes
of differently hot parts of the system (defined as the intercept 
of a particular isotherm with the $t$--axis) as a function of
the initial energy density. 
Assuming that the system freezes out at these temperatures, 
we then show how the lifetimes can be
inferred from two--particle correlation functions.

\subsection{Lifetimes}

For convenience, we first show in Fig.\ 10 the lifetimes as a function
of $\epsilon_0$ for the one--dimensional (Landau) expansion 
studied in \cite{dhrmg}. Figs.\ 10 (a) and (b) are for
the case $d_Q/d_H=37/3$, (c) and (d) are for $d_Q/d_H=3$.
The thick lines in (a,c) show the lifetimes of matter with
$T=0.7\, T_c$ (solid line), $T= 0.9\, T_c$ (dotted), 
and $T=T_c$ (dashed) for $\Delta T=0$, the
corresponding lines in (b,d) are for finite $\Delta T=0.1\, T_c$.
For comparison, the thin lines are the corresponding lifetimes
for the expansion of an ideal gas with $d_H$ degrees of freedom
(they are identical in (a) and (b) as well as in (c) and (d)).

One clearly observes the maximum in the lifetime of mixed phase
matter ($T=T_c$) at $\epsilon_0=\epsilon_Q$ in Figs.\ 10 (a,c).
As discussed in \cite{dhrmg}, this maximum vanishes for
finite $\Delta T$, Figs.\ 10 (b,d). Moreover, such hot matter does
not survive as long as in the ideal gas case. The lifetime of cooler matter
with $T=0.7\, T_c$, however, stays longer than in the expansion
of an ideal gas, independent of the value of $\Delta T$. Note the change
in the scale of $t_{\rm \,life}$ by a factor of 3 between (a,b) and (c,d).
This is due to the reduction of the latent heat for a smaller ratio
$d_Q/d_H$, which in turn accelerates the expansion.

Fig.\ 11 shows the corresponding diagram for the spherical expansion.
First of all, one notices that, at high $\epsilon_0$, 
there is almost no difference between the lifetime of a system with
a transition to the QGP and an ideal gas. Inspecting Fig.\ 5, one
observes, however, that the bulk of matter at finite $r$
does indeed live longer in the first case. 
Thus, for the spherical expansion, our
definition of `lifetime' should rather be replaced by an average 
over the particular isotherm. It is, however, not necessary
to do so at this point, since the correlation 
functions considered subsequently will take this into account
in a natural way.

Second, one observes that the lifetimes do not grow as strongly
for high $\epsilon_0$ as in the
one--dimensional expansion. This is due to the
fact that the system disperses its initial internal energy
much more efficiently into kinetic energy in three
dimensions than it does in one dimension. This also leads
to the reduction of the overall scale in the lifetimes as
compared to Fig.\ 10.

Third, it is noticeable that the {\em increase\/} of the lifetime
for $\Delta T=0$ at $\epsilon_0 = \epsilon_Q$ (where the lifetime is 
maximum) as compared to the ideal gas case is about a 
factor of 2 bigger than in the one--dimensional expansion. The
prolongation of the lifetime is thus most pronounced in spherical
geometry. Unfortunately, the reduction of the lifetime
in the case of a smooth transition is also rather strong, cf.\
Figs.\ 11 (b,d). Nevertheless, for freeze-out at $T=0.7\, T_c$
the lifetime can still be longer by a factor of 2 in the case
of a transition than in the expansion of an ideal gas.
Moreover, in contrast to the one--dimensional case,
the spherical geometry leads to a (broad) maximum in the lifetime
around $\epsilon_Q$.

In Fig.\ 12 we present the lifetimes for the expansion of a
Bjorken cylinder with initial time $\tau_0=0.1\, R$. One
still observes the distinguished maximum in the lifetime
associated with the transition to the QGP as seen in
the preceding Fig.\ 11. In this case, however,
the overall scale is even smaller (due to the fact that
the system is from the very beginning quite effectively
diluted by the longitudinal velocity field).

Moreover, one observes a shift in the maximum of the
lifetime. This was discussed in the preceding section and is
due to the fact that the longitudinal motion 
compensates the tendency to explode
transversally (on account of a high initial energy density).
For $\Delta T=0$ the maximum is now around 
$\epsilon_0 \sim 20 - 50\, T_c\, s_c$
for {\em both\/} values of $d_Q/d_H$.
For $\Delta T=0.1\, T_c$, one still observes a maximum
in the lifetimes, but as in Fig.\ 11 it is also broader and less
pronounced.

The corresponding plot for $\tau_0 = 0.5\, R$ looks rather similar,
and will therefore not be shown here. The position of the maximum is,
however, shifted to smaller $\epsilon_0 \sim 3-7\, T_c\, s_c$, for
reasons discussed above. The situation for a dynamical
$\tau_0 = 1/3\, T_0$ is shown in Fig.\ 13. As one expects from the
above discussion, there is a steady increase of the lifetime
up to the highest considered $\epsilon_0$.

\subsection{Two--particle correlation functions}

To calculate two--particle correlation functions we use the method
developped by Pratt \cite{pratt}, Sinyukov \cite{sinyukov} and others,
and applied to hydrodynamics by the Marburg group \cite{schlei}.
This method is essentially a straightforward generalization
of the Cooper--Frye formula \cite{frye} for single inclusive
particle spectra to the case of correlation functions. 

To calculate single inclusive particle spectra or two--particle
correlation functions in the hydrodynamical framework, one assumes
that a fluid element decouples from the fluid evolution (``freezes
out'') as soon as
the particle density drops below a certain critical value where
the collision rate becomes too small to maintain (local) thermodynamical
equilibrium. In our case, the particle density depends on the
temperature $T$ only. Therefore, freeze-out happens across
isotherms in the space--time diagram. To incorporate the freeze-out
consistently into the solution of the hydrodynamical equations is
a non-trivial problem. 
Hydrodynamics is obviously not applicable to determine the motion
of particles that are already frozen out. Therefore, the hydrodynamical
solution has to be restricted to a limited region of space--time.
The main difficulty is that its boundary, i.e., the freeze-out
hypersurface, has to be determined {\em dynamically}, i.e.,
simultaneously with the solution of the hydrodynamical equations. A possible
treatment of this problem was recently proposed by Bugaev \cite{bugaev}, 
but has so far not been applied in practical calculations.

Commonly one circumvents this problem assuming
the validity of the hydrodynamical description in the {\em whole\/}
forward light cone, then solves the hydrodynamical equations, and  
determines the isotherms. Finally, one employs the
Cooper--Frye formula \cite{frye} to calculate particle
spectra along the isotherm corresponding to the particular
freeze-out temperature (for a detailed discussion of this
approach see, for instance, \cite{bernard}). 
We note that the Cooper--Frye formalism holds rigorously only on
space-like hypersurfaces. For time-like hypersurfaces 
it may yield negative numbers of particles emitted from
the hypersurface (corresponding to particles that do not freeze out
but reenter the fluid). To cure this obviously
unphysical result, modifications have been proposed in \cite{bugaev,sin2}.
The approach of Ref.\ \cite{sin2} is, however, problematic
since it does not describe correctly radiation from a static source.
When modifying the particle spectra as proposed in \cite{bugaev}
we found, however, for the cases under consideration
virtually no deviation to the results obtained with the
conventional approach of Cooper and Frye. Therefore, we will use
this well-established method throughout the following.

The two--particle correlation function measures the coincidence probability
$P({\bf p}_1, {\bf p}_2)$ of two (identical) particles with 
momenta ${\bf p}_1,\, {\bf p}_2$ relative
to the probability of detecting uncorrelated particles
from different events,
\begin{equation}
C_2 ({\bf p}_1, {\bf p}_2) = \frac{P({\bf p}_1, {\bf p}_2)}{P({\bf p}_1)
P({\bf p}_2)}\,\, .
\end{equation}
In the following, the average 4--momentum is denoted as $K^{\mu} = (p_1^{\mu}
+p_2^{\mu})/2$ and the relative 4--momentum as $q^{\mu}=p_1^{\mu} - p_2^{\mu}$.
Under the assumption that the particle 
source is chaotic and sufficiently large,
and that the emitted particles are bosons (with degeneracy factor $d$)
the two--particle correlation function can be written as \cite{schlei}
\begin{equation} \label{c2}
C_2 ({\bf p}_1, {\bf p}_2) = 1 + \frac{\left| \frac{d}{(2\pi)^3}
\int_{\Sigma}\, {\rm d}
\Sigma \cdot K\,\, \exp\, [i\, \Sigma \cdot q] \,\, 
f\left(\frac{u \cdot K}{T}\right)
\right|^2}{ E_1 \, \frac{{\rm d}N}{{\rm d}^3 {\bf p}_1} \,\,\, E_2\,
\frac{ {\rm d}N}{{\rm d}^3 {\bf p}_2} }\,\, ,
\end{equation}
where \cite{frye}
\begin{equation} \label{single}
E\, \frac{{\rm d}N}{{\rm d}^3 {\bf p}} = \frac{d}{(2\pi)^3}
\int_{\Sigma}\, {\rm d}
\Sigma \cdot p \,\, f\left( \frac{u \cdot p}{T}\right)
\end{equation}
is the single inclusive momentum distribution. $f(x)=(e^x-1)^{-1}$ 
is the Bose--Einstein distribution function, and
$u^{\mu}$ the fluid 4--velocity. The integrals
run over the freeze-out hypersurface. In general, that hypersurface is 
represented by a 3--parametric (4--vector) function 
$\Sigma^{\mu}(\zeta,\eta,\phi)$, and
the normal vector on the hypersurface is determined by
\begin{equation} \label{normal}
{\rm d} \Sigma_{\mu} = \epsilon_{\mu \alpha \beta \gamma}\,
\frac{\partial \Sigma^{\alpha}}{\partial \zeta}\, 
\frac{\partial \Sigma^{\beta}}{\partial \eta}\, 
\frac{\partial \Sigma^{\gamma}}{\partial \phi}\, 
{\rm d}\zeta\, {\rm d}\eta\, {\rm d}\phi\,\, ,
\end{equation}
where $\epsilon_{\mu \alpha \beta \gamma}$ ($=-1$ for $(\mu \alpha
\beta \gamma)$ an even permutation of $(0\,1\,2\,3)$) is the completely
antisymmetric 4--tensor. While the (freeze-out)
temperature $T$ is (by definition) assumed to be 
constant along the freeze-out hypersurface, 
the fluid velocity varies, $u^{\mu} = u^{\mu}(\Sigma)$.

For symmetric systems, the number of independent variables
which $C_2$ depends on can be reduced.
For the spherically symmetric fireball geometry, the orientation
of the average momentum can be chosen arbitrarily, such that $C_2$ 
depends only on the modulus of ${\bf K}$. Our choice of coordinate
system will be such that ${\bf K} = (0,0,K)$.
Furthermore, ${\bf q}$ can be decomposed into a so-called
``out'' component ${\bf q}_{\rm \,out} = (0,0,q_{\rm \,out})$ 
parallel to ${\bf K}$ and a 
``side'' component ${\bf q}_{\rm \,side}$ orthogonal to ${\bf K}$ and 
${\bf q}_{\rm \,out}$. Rotational symmetry around the direction of ${\bf K}$
allows us to choose ${\bf q}_{\rm \,side} = (q_{\rm \,side},0,0)$, such
that the correlation function depends only on three independent variables,
$C_2(K,\, q_{\rm \,out},\, q_{\rm \,side})$.

For the Bjorken cylinder geometry, we restrict our consideration to
particles emitted at midrapidity, $K^z=q^z=0$. Rotational symmetry
around the $z$--axis allows us to choose the average
transverse momentum as ${\bf K}_\perp = (K,0,0)$, and consequently,
${\bf q}_{\rm \,out}=(q_{\rm \,out},0,0)$, ${\bf q}_{\rm \,side} = 
(0,q_{\rm \,side},0)$. Again, $C_2(K,q_{\rm \,out},q_{\rm \,side})$ is
a function of three independent variables only.
The explicit evaluation of (\ref{c2})
for the fireball and Bjorken cylinder geometry is referred to
Appendices A and B, respectively.

As is well known \cite{pratt}, the width of the correlation
function in out--direction is inversely proportional to the duration
of particle emission, i.e., to the lifetime of the source.
Analogously, the inverse width of the correlation function in
side--direction is a measure for the (transverse) size of the source.
To be more precise, for fixed average (transverse) momentum $K$
we define side-- and out--correlation functions as
$C_{2,{\rm \,side}}(q_{\rm \,side}) \equiv C_2(K,0,q_{\rm \,side})$
and $C_{2,{\rm \,out}}(q_{\rm \,out}) \equiv C_2(K,q_{\rm \,out},0)$,
respectively.
We furthermore define a corresponding inverse width as
$R_{\rm \,side} \equiv 1/q_{\rm \,side}^*$, where
$q_{\rm \,side}^*$ is determined by $C_{2,{\rm \,side}}(q_{\rm
\,side}^*) =1.5$, and analogously for $R_{\rm \,out}$.

We emphasize that we do not attempt a standard Gaussian
fit of the two--particle correlation function, which would
relate the inverse widths to the usual radii parameters.
First of all, we will be interested only in the generic shape
of the correlation functions which is satisfactorily
characterized by their inverse widths $R_{\rm \,side},\, R_{\rm \,out}$.
Second, the functional form of the correlation functions is 
not a Gaussian (cf.\ the explicit formulae in Appendix A and B). 
The fit procedure would therefore only introduce unnecessary
errors. Third, the radii parameters as well as the inverse widths 
are only {\em proportional\/}
to the actual (average) size and lifetime of the system.
It is well known that hydrodynamical flow affects the radii
parameters/widths, and, moreover,
that this effect is sensitive to the choice of $K$ \cite{pratt,schlei}. 
It is therefore tedious (if not impossible) to relate $R_{\rm \,side}$ 
and $R_{\rm \,out}$ to the real source size and
lifetime. This holds for our model calculations as well as for 
the experiment.

We can expect, however, that such effects are either irrelevant or
largely cancel out
if we consider the {\rm ratio\/} $R_{\rm \,out}/R_{\rm \,side}$.
As can be inferred from the space--time diagrams in Figs.\ 3--5 and
7--9, while the size of the system is approximately constant, it is the
lifetime that varies appreciably, depending on whether the system undergoes
a phase transition or not. Thus, $R_{\rm \,out}/R_{\rm \,side}$
can be expected to be a good measure for the lifetime of the system.
In the following, we choose the particles to be pions with mass
$m=138$ MeV, and as average momentum $K=300$ MeV. On one
hand, this value is well within the typical experimental acceptance.
On the other hand, it is large enough to exhibit effects of the
prolonged lifetime in the case of a
transition to the QGP \cite{pratt}. To fix the $q$--scale in MeV
we take $R=5$ fm as a typical initial radius of the system.

In Fig.\ 14 we show as an example
side-- and out--correlation functions corresponding
to the hydrodynamical evolution of Fig.\ 4. In light of the 
uncertainties in the actual freeze-out temperature, we calculate the
correlation functions along isotherms for $T=T_c, \, 0.9\, T_c$, as well
as $0.7\, T_c$. First of all, one immediately recognizes that in all cases
the widths of the correlation functions correspond closely
to the space--time structure of the corresponding isotherms. 
For instance, the exceedingly long lifetime of the system in Fig.\ 4 (b)
is reflected in the comparatively small width of the out--correlation function
Fig.\ 14 (b). On the other hand, the transverse size is about the same
in all cases, cf.\ Figs.\ 4 (b,d,f), which reflects in nearly 
identical side--correlation functions in Figs.\ 14 (a,c,e).
Furthermore, with the exception of the case $\Delta T=0.1\, T_c$,
the correlation functions are almost completely
insensitive to the freeze-out temperature chosen. This is intuitively
clear since the corresponding space--time isotherms differ only marginally
for $\Delta T=0$ and the ideal gas calculation, cf.\ Figs.\ 4 (b,f), while
there are larger differences in Fig.\ 4 (d).
Note that the results shown in Figs.\ 14 (a,b,e,f) correspond nicely 
to those in Fig.\ 4 of \cite{pratt}, in spite of the fact that
the latter calculation employs inhomogeneous initial conditions, a
different way to solve the hydrodynamical equations, and a different
treatment of the freeze-out\footnote{The freeze-out in \cite{pratt}
is in some sense performed dynamically, the freeze-out surface, however, 
is assumed to have no time-like parts.}.
In Fig.\ 15 we show the correlation functions corresponding to the
hydrodynamical evolution of Fig.\ 9. Again, they
adequately characterize the space--time geometry of the source.

Fig.\ 16 shows the experimentally measurable
ratio $R_{\rm \,out}/R_{\rm \,side}$ as a function of
$\epsilon_0$ for the spherical fireball geometry. Comparing the results
with Fig.\ 11, one observes that this ratio reflects
closely the behaviour of the lifetime of the system, independent of details
in the equation of state such as the width of the transition region
$\Delta T$ or the latent heat of the transition (which is
proportional to $d_Q/d_H$). Also, for the case of a first order
transition, $\Delta T=0$, Figs.\ 16 (a,c), the enhancement in 
$R_{\rm \,out}/R_{\rm \,side}$ over the ideal gas case is a factor
of 3 to 7 (for $d_Q/d_H= 3$ to $37/3$) at $\epsilon_0 \sim \epsilon_Q$.
In the case of a smooth transition, $\Delta T=0.1\, T_c$, Figs.\
16 (b,d), this is considerably reduced (as expected from Fig.\ 11), but
if the system freezes out at temperatures $T_f \leq0.7\, T_c$, there is
still a factor of 2 enhancement over the ideal gas case.

In Figs.\ 17, 18 we present the corresponding results for the Bjorken cylinder
expansion with $\tau_0 = 0.1\, R$ and $\tau_0=1/3\, T_0$, respectively.
In all cases we find that the experimentally measurable
ratio of correlation widths mirrors closely the dependence of the lifetime
on initial conditions in Figs.\ 12, 13. 
The most favourable case $R_{\rm \,out}/R_{\rm \,side} \sim 3.5$ (for
a strong first order transition with a large latent heat, $d_Q/d_H=37/3$, 
cf.\ Fig.\ 17 (a)), may be reached with initial conditions expected at 
RHIC energies. At that point the enhancement of this ratio
is about a factor of 2 above the ideal gas case and virtually
independent of the freeze-out temperature.
In the scenario with $\tau_0=1/3\, T_0$, Fig.\ 18 (a), 
the enhancement is somewhat smaller and is shifted toward higher
initial energy densities because the initial pure Bjorken expansion phase
starts earlier than in the fixed $\tau_0=0.1\, R$ case.
For a smooth transition with $\Delta T=0.1\, T_c$, 
the maximum ratio is reduced to 2.5 and varies 
less rapidly with $\epsilon_0$, but is still about $40\%$ larger 
than for the ideal gas expansion. In this case, however,
a significant time delay can only be observed if 
the freeze-out occurs relatively late with $T_f \sim 0.7\, T_c$.
As seen in Figs.\ 17, 18, 
for earlier freeze-out, smaller $\tau_0$, or smaller $d_Q/d_H$
the enhancement relative to the ideal gas case is significantly reduced
and would be more difficult to observe.

At energy densities
estimated to be reached in CERN SPS Pb+Pb--collisions ($\epsilon_0 \sim
1-2\, T_c\, s_c$ in our units), we expect from our results
that $R_{\rm \,out}/R_{\rm \,side} \sim 1.5 - 2$. 
However, present data from CERN SPS \cite{na44} 
indicate that the (fitted) out--radii are rather similar 
to the side--radii. This does not contradict our results, because, as 
shown by Schlei et al.\ \cite{schlei2} in the framework of a hydrodynamical
calculation similar to ours, 
correlation functions constructed from thermal pions only
give $R_{\rm \,out}/R_{\rm \,side} \sim 2$ (cf.\ especially 
\cite{schlei}), while the incorporation of long-lived
{\em resonance decays\/} leads to a reduction of that ratio and
good agreement with the measured radii.
We note that kaon interferometry \cite{miklos,schlei3,future}
is preferable, though experimentally more difficult,
because only distortions of the interference pattern
due to shorter lived $K^*$ resonances have to be taken into account.

\section{Conclusions}

In this paper we investigated the spherically symmetric expansion
of a fireball as well as the cylindrically symmetric transverse
expansion of a QGP with boost-invariant initial conditions
along the beam axis. The expansion was treated in the framework of
ideal relativistic hydrodynamics and extends our systematic
study \cite{test1,dhrmg,puersuen,test2,bernard}
of collective flow patterns with realistic equations of state.
The symmetries of the considered
geometries allowed us to calculate the flow patterns
using a simple modification of a well-tested one--dimensional algorithm.

The emphasis of the present investigation was on how 
a rapid cross-over to the QGP in the equation of state 
influences the collective expansion dynamics.
In particular, since present lattice data only constrain the width of
the transition region to the QGP to be in the range 
$0\leq \Delta T < 0.1\, T_c$, it is important
to test how such uncertainties may influence dynamical observables.
We also studied the effect of varying the ratio of degrees
of freedom in the QGP and hadron phase (i.e.\ essentially the latent
heat of the transition) on the system dynamics. The results were
compared to the expansion of an ultrarelativistic 
ideal gas (without transition) using fixed energy density initial conditions.

We focussed on the lifetime of the system as a function of
initial energy density as an important {\em collective\/} observable 
that can discriminate between different equations of state.
As expected from previous one--dimensional studies in the framework of
the Landau expansion model \cite{dhrmg}, 
we found that the lifetime of a spherical fireball is much longer
in the case of a first order phase transition, $\Delta T=0$,
as compared to the expansion of an ideal gas without transition. The
prolongation of the lifetime in that case can be up to a factor of 4.5 to 9
(for $d_Q/d_H=3$ to $37/3$, respectively),
provided the initial energy density corresponds to that of mixed phase
with a large fraction of QGP. This time delay
effect was originally pointed out by Pratt
\cite{pratt}. In the case of a smooth transition, $\Delta T= 0.1\, T_c$,
however, this time delay is drastically reduced. The lifetimes 
(of matter with $T=0.7\, T_c$) nevertheless still remain on the order of
a factor of 2 longer as compared to the ideal gas expansion.

For the Bjorken cylinder expansion, it is
also necessary to specify the initial
(proper) time $\tau_0$ for the onset of hydrodynamic expansion. 
To explore uncertainties
associated with this additional degree of freedom, we investigated the cases
$\tau_0= 0.5\, R,\, 0.1\, R$, and a dynamical $\tau_0=1/3\,T_0$ varying
with the initial temperature. The results were similar as for the 
spherical expansion, up to two important exceptions:
(a) the maximum lifetimes emerged at higher initial energy densities
(the exact value of which depends on the choice of $\tau_0$)
corresponding to QGP matter instead of mixed phase matter, 
and (b) the lifetimes were in general shorter.

Both effects are explained by the very efficient cooling due to the
initial longitudinal velocity profile associated with the boost
invariance of the problem. This effect causes an overall
reduction of the lifetimes. Moreover, in order for 
slow (van Hove \cite{vanH}) 
deflagration fronts to dominate the cooling mechanism 
in transverse direction, one has to start at higher initial energy 
densities to compensate for the longitudinal cooling. Otherwise, 
the longitudinal cooling reduces the energy densities too fast and 
the associated deflagration solution \cite{test1} vanishes. Further cooling
is then achieved (in the usual way) through a (fast) simple wave.

Finally, we showed (cf.\ also \cite{pratt,bertsch})
that the prolongation of the lifetime can be
observable via the ratio $R_{\rm \,out}/R_{\rm \,side}$ of inverse
widths of two--particle correlation functions in out-- and side--direction.
This ratio follows the behaviour of the lifetimes rather closely. The 
prolongation of the lifetime in the case of a transition to the
QGP could therefore be in principle searched for using 
this observable. The enhancement of that ratio
is, of course, strongest in the case that the transition is first order with
a large latent heat. An interesting result is that, for the fireball geometry, 
the effect is maximum for energy densities 
achieved at the AGS, while for the Bjorken cylinder geometry,
the maximum of $R_{\rm \,out}/R_{\rm \,side}$ occurs at energy densities
presumably reached at the RHIC collider. 

There are several effects which may reduce the strength of the time--delay
signal observable via the $R_{\rm \,out}/R_{\rm \,side}$--ratios
that will require further investigation.
First, for the fireball geometry 
it is important to extend our considerations to finite baryon number density.
At finite chemical potential the width of the cross-over region
may be significantly larger than at zero chemical potential.
Second, the decay of long-lived resonances can simulate
time delay \cite{padula}.
Interferometry with kaons instead of pions is therefore preferable
\cite{future}. Finally, while our investigations covered a wide
range of uncertainties
in the equation of state, our calculations have neglected effects of
dissipation that tend in general to reduce
the collective flow strengths predicted via ideal hydrodynamics.
For instance, bulk viscosity appears
in the hydrodynamical equations of motion in a similar way as the
pressure, and could in principle counteract any reduction of the
velocity of sound in the transition region. 
The main result of this paper is, nevertheless, that the generic
time--delay signature of QGP formation is remarkable robust to present
uncertainties in the QCD equation of state.
\\ ~~ \\
\noindent 
{\bf Acknowledgments}
\\ ~~ \\ 
We thank G.\ Bertsch, A.\ Dumitru, and E.\ Shuryak for stimulating
discussions on the lifetime of the QGP. 
One of us (D.H.R.) benefitted considerably
from discussions with B.\ Schlei on the 
calculation of two--particle correlation functions
in the hydrodynamical framework. We thank U.\ Heinz and W.\ Zajc
for useful discussions on our results 
and D.\ Keane for emphasizing that kaons might be the ideal
probe to detect a prolonged lifetime with the STAR detector at RHIC.
\newpage
\appendix

\section*{Appendix A}
In this Appendix we explicitly calculate the out-- and side--correlation
functions in the fireball geometry. We first specify
the parameters of the freeze-out hypersurface. For the following discussion,
we conveniently choose spherical coordinates 
$r,\, \theta,\, \varphi$ in space. Let
(a) $\zeta$ parametrize the surface in the $t-r$ plane, 
$0 \leq \zeta \leq 1$, with the lower boundary corresponding 
to $t=0$ and the upper to $r=0$. Let (b) $\eta$ parametrize the 
surface in the $r-\theta$ plane and (c) $\phi$ in the $r-\varphi$ plane.
In the $t-r$ plane, the surface is given by the corresponding isotherm
shown in Figs.\ 3--5. Due to spherical symmetry, however, the surface
in the $r-\theta$ and $r-\varphi$ planes is trivial. We can simply
identify $\eta=\theta$, $\phi= \varphi$ and vary them
within the standard boundaries for the azimuthal and polar angle. As a
consequence, the time $t_f$ and the radius $r_f$
at freeze-out depend only on $\zeta$, not on $\eta$ or $\phi$.
The freeze-out surface is thus given by $\Sigma^{\mu} = 
(t_f(\zeta),r_f(\zeta)\, {\bf e}_r)$, where ${\bf e}_r =
(\sin \eta\, \cos \phi, \sin \eta\, \sin \phi, \cos \eta)$.

Applying (\ref{normal}) and choosing the sign to have
${\rm d} \Sigma_{\mu}$ pointing outwards along the isotherm, 
we obtain
\begin{equation} \label{dsigma}
{\rm d}\Sigma_{\mu} = \left(-\frac{{\rm d}r_f}{{\rm d}\zeta},
\frac{{\rm d}t_f}{{\rm d}\zeta}\, {\bf e}_r \right) \, r_f^2(\zeta)\,
\sin \eta \,\, {\rm d}\zeta\, {\rm d}\eta \, {\rm d}\phi\,\, .
\end{equation}
For the calculation of the correlation function we employ the
Boltzmann approximation. For an average momentum $K=300$ MeV chosen
in our calculations the error introduced is negligible, in particular
since we also consider the particles to
be massive pions, $m = 138$ MeV, and choose $T_c=160$ MeV.

For the single inclusive pion spectrum (\ref{single}), one
conveniently chooses ${\bf p} = (0,0,p)$, and
obtains after inserting (\ref{dsigma}) and performing the $\phi$--
and $\eta$--integrations
\begin{equation} \label{single2}
E\, \frac{{\rm d}N}{{\rm d}^3 {\bf p}} = 
\frac{d}{2 \pi^2} \int_0^1 {\rm d}\zeta \, r_f^2(\zeta)\, e^{-E\gamma /T}
\left\{ - E\, \frac{\sinh a}{a}\,\, \frac{{\rm d}r_f}{{\rm d} \zeta}
        + p\, \left[ \frac{\cosh a}{a} - \frac{\sinh a}{a^2} \right]\,
       \frac{{\rm d}t_f}{{\rm d}\zeta} \right\} \, \,  ,
\end{equation}
where $a\equiv p v \gamma /T$, 
$E=[{\bf p}^2 +m^2]^{1/2}$, and $v$ is the (radial) fluid 3--velocity,
$\gamma=[1-v^2]^{-1/2}$. Here (and in the following), the
$\zeta$--integration has to be done numerically along the
respective isotherms obtained from the solution of the
hydrodynamical equations. For further use, we define
\begin{equation} \label{i0}
{\cal I}_0 (p) \equiv \frac{(2\pi)^3}{d}\, E \, 
\frac{{\rm d}N}{{\rm d}^3 {\bf p}} \,\, .
\end{equation}
For the side--correlation function, $q_{\rm \,out} =0$, and consequently
\begin{eqnarray}
p_1^{\mu} & = & ( E,q_{\rm \,side}/2,0,K)\,\, , \\
p_2^{\mu} & = & ( E,-q_{\rm \,side}/2,0,K)\,\, , \\
K^{\mu} & = & ( E,0,0,K)\,\, , \\
q^{\mu} & = & (0,q_{\rm \,side},0,0)\,\, , 
\end{eqnarray}
where $E=[K^2 + q_{\rm \,side}^2/4 + m^2]^{1/2}$.
Since the single inclusive spectrum (\ref{single2}) does not depend on
the direction of ${\bf p}$, the two single inclusive 
spectra in the denominator in (\ref{c2})
are equal. Furthermore, for the numerator one has to calculate the
expressions
\begin{eqnarray}
{\cal I}_1 & \equiv & 
\int_{\Sigma} {\rm d}\Sigma \cdot K \, e^{-u \cdot K /T} \,
\cos\, [ \Sigma \cdot q ]\,\, , \\
{\cal I}_2 & \equiv &
\int_{\Sigma} {\rm d}\Sigma \cdot K \, e^{-u \cdot K /T} \,
\sin\, [ \Sigma \cdot q ]\,\, .
\end{eqnarray}
For the calculation of ${\cal I}_1$ we insert (\ref{dsigma}) and perform the 
$\phi$--integration with the help of eq.\ (3.715.18) of Ref.\ \cite{GR}. 
Subsequently, the
$\eta$--integration can be done with an analytic continuation of
either eq.\ (6.616.5) or (6.677.6) of \cite{GR} and a suitable first
derivative of these formulae. The final result reads
\begin{eqnarray} \label{i1}
{\cal I}_1 & = & 4 \pi \int_0^1 {\rm d}\zeta \, r_f^2(\zeta)\, 
e^{-E\gamma /T} \\
& \times & \left\{ - E\, \, \frac{ \sinh \sqrt{a^2-b^2}}{\sqrt{a^2-b^2}} \,
\frac{{\rm d}r_f}{{\rm d}\zeta} + K\, \frac{a}{\sqrt{a^2-b^2}}\,
\left[ \frac{\cosh \sqrt{a^2-b^2}}{\sqrt{a^2-b^2}} - \frac{\sinh
\sqrt{a^2-b^2}}{a^2-b^2} \right] \, \frac{{\rm d}t_f}{{\rm d}\zeta} \right\}
\,\, , \nonumber
\end{eqnarray}
where $a \equiv Kv\gamma /T$, $b\equiv q_{\rm \,side}\, r_f$. It is easy
to show that for the side--correlation function, ${\cal I}_2 \equiv 0$ by
symmetry. The final result thus reads
\begin{equation} \label{c2side}
C_{2,{\rm \, side}} = 1 + {\cal I}_1^2/{\cal I}_0^2\,\, ,
\end{equation}
with ${\cal I}_0$ from (\ref{i0}) and ${\cal I}_1$ from (\ref{i1}).

For the out--correlation function, $q_{\rm \,side} =0$, and consequently
\begin{eqnarray}
p_1^{\mu} & = & ( E_1,0,0,K+q_{\rm \, out}/2)\,\, , \\
p_2^{\mu} & = & ( E_2,0,0,K-q_{\rm \, out}/2)\,\, , \\
K^{\mu} & = & ( K^0,0,0,K)\,\, , \\
q^{\mu} & = & (E_1-E_2,0,0,q_{\rm \,out})\,\, ,
\end{eqnarray}
where $E_{1,2}=[(K \pm q_{\rm \,out}/2)^2 + m^2]^{1/2},\,K^0 = (E_1+E_2)/2$. 
The $\phi$--integration in the calculation of ${\cal I}_1,\,
{\cal I}_2$ is now trivial. 
For the $\eta$--integration, we employ the angle addition
theorem and eqs.\ (2.662.1,2) of \cite{GR}, obtaining
\begin{eqnarray}
{\cal I}_1 & = & 4 \pi \int_0^1 {\rm d}\zeta\, r_f^2(\zeta)\, 
e^{-K^0\gamma/T}
\left\{- K^0\, \left[ \cos \beta \, {\cal J}_0(a,b) + \sin \beta \,
{\cal J}_1(a,b) \right] \,\frac{{\rm d}r_f}{{\rm d}\zeta} \right. \nonumber
\\
&   & \hspace*{4cm} \left. + K \left[
\cos \beta \, \frac{\partial {\cal J}_0(a,b)}{\partial a} +\sin \beta \,
\frac{\partial {\cal J}_1(a,b)}{\partial a} \right] \,
\frac{{\rm d}t_f}{{\rm d}\zeta}
\right\} \, \, ,
\end{eqnarray}
where $\beta \equiv (E_1-E_2)t_f$, and
\begin{equation}
{\cal J}_0 (a,b) \equiv \frac{ a\, \cos b \, \sinh a + b\, \sin b \, \cosh a}{
a^2 + b^2}\,\, , \,\,\,\, {\cal J}_1(a,b) \equiv \frac{ a\, \sin b \, \cosh a -
 b\, \cos b \, \sinh a}{ a^2 + b^2}\,\, ,
\end{equation}
with $a \equiv Kv\gamma/T$ and $b\equiv q_{\rm \,out}\, r_f$.
In complete analogy one obtains ${\cal I}_2$ as
\begin{eqnarray}
{\cal I}_2 & = & 4 \pi \int_0^1 {\rm d}\zeta\, r_f^2(\zeta)\, 
e^{-K^0\gamma/T}
\left\{- K^0\, \left[ \sin \beta \, {\cal J}_0(a,b) - \cos \beta \,
{\cal J}_1(a,b) \right] \,\frac{{\rm d}r_f}{{\rm d}\zeta} \right. \nonumber
\\
&   & \hspace*{4cm} \left. + K \left[
\sin \beta \, \frac{\partial {\cal J}_0(a,b)}{\partial a} - \cos \beta \,
\frac{\partial {\cal J}_1(a,b)}{\partial a} \right] \,
\frac{{\rm d}t_f}{{\rm d}\zeta}
\right\} \, \, .
\end{eqnarray}
Finally, the out--correlation function reads
\begin{equation} \label{c2out}
C_{2,{\rm \,out}} = 1 + \frac{{\cal I}_1^2 + {\cal I}_2^2}{{\cal I}_0(p_1)
\,\,{\cal I}_0(p_2)}\,\, .
\end{equation}

\section*{Appendix B}
In this Appendix we explicitly calculate the out-- and side--correlation
functions in the Bjorken cylinder geometry. Convenient coordinates
to work in are cylindrical coordinates $r,\, \varphi,\, z$ in space. 
As in the preceding Appendix, (a) $\zeta$ parametrizes the freeze-out 
surface in the $t-r$ plane (at $z=0$), $0\leq \zeta \leq 1$, 
$t_f(0)=\tau_0,\, r_f(1)=0$.
(b) $\eta$ parametrizes the surface in the $t-z$ plane. Due to boost
invariance, that surface is simply the space--time hyperbola of constant
$\tau_f = \sqrt{t_f^2 - z_f^2}$. Note that at $z=0$,
$\tau_f = t_f(\zeta)$ and therefore $\tau_f$ depends on $\zeta$. 
A natural choice for the parameter $\eta$ 
is the space--time rapidity Artanh$\,[z_f/t_f]$ (as
implied by our choice of symbols), such that
$t_f (\zeta,\eta)= \tau_f(\zeta)\, \cosh \eta$, 
$z_f(\zeta,\eta)= \tau_f (\zeta)\, \sinh \eta$.
Finally, (c) $\phi$ parametrizes the hypersurface in the $r-\varphi$ plane
and due to cylindrical symmetry can be identified with the polar angle,
$\phi \equiv \varphi$.
We note that due to boost invariance along $z$, $r_f$ cannot depend
on $\eta$.

The hypersurface 4--vector thus reads 
\begin{equation}
\Sigma^{\mu} = \left (\tau_f(\zeta)\, \cosh \eta, \, r_f(\zeta)\, \cos \phi,
\, r_f (\zeta) \, \sin \phi, \, \tau_f(\zeta)\, \sinh \eta \right)\,\, ,
\end{equation}
with $\tau_f(\zeta) \equiv t_f(\zeta,\eta=0)$, i.e., given
by the isotherms in Figs.\ 7--9. The normal vector (with
proper orientation) is readily calculated with (\ref{normal}),
\begin{equation}
{\rm d}\Sigma_{\mu} = \left( -\frac{{\rm d}r_f}{{\rm d}\zeta}\, \cosh
\eta, \, \frac{{\rm d}\tau_f}{{\rm d}\zeta}\, \cos \phi,\, 
\frac{{\rm d}\tau_f}{
{\rm d}\zeta}\, \sin \phi, \, \frac{{\rm d}r_f}{{\rm d}\zeta}\, \sinh \eta
\right)\, r_f(\zeta)\, \tau_f(\zeta)\, \, {\rm d} \zeta\, {\rm d} \eta\,
{\rm d} \phi\,\, .
\end{equation}
For the calculation of the single inclusive spectrum, we employ
$m_\perp \equiv [{\bf p}_\perp^2 + m^2]^{1/2}$, i.e., $p^0 = m_\perp\, 
\cosh y, \,
p^z = m_\perp\, \sinh y$, with the longitudinal (particle) rapidity
$y \equiv {\rm Artanh}\, [p^z/E]$. Furthermore, due to rotational
symmetry around $z$, we may choose ${\bf p}_\perp = (p_\perp,0,0)$. Thus,
\begin{equation} \label{dsigmadotp}
{\rm d} \Sigma \cdot p = \left(- m_\perp \cosh \,[y-\eta] \, \frac{{\rm d}
r_f}{{\rm d} \zeta} + p_\perp  \cos \phi \,\, \frac{{\rm d}
\tau_f}{{\rm d} \zeta} \right) r_f(\zeta) \, \tau_f (\zeta) \, {\rm d}
\zeta \, {\rm d} \eta\, {\rm d} \phi\,\,.
\end{equation}
The fluid 4--velocity in the Bjorken cylinder expansion reads
$u^{\mu} = \gamma\, (1,\,v \,{\bf e}_r,\,z/t)$, where
$v\equiv v_\perp,\, {\bf e}_r = (\cos \phi, \sin \phi,0)$. 
With the (longitudinal)
space--time rapidity $\eta \equiv {\rm Artanh}\, [z/t]$ and the
transverse fluid rapidity $\eta_r \equiv {\rm Artanh}\, [v\, \cosh
\eta]$, one obtains
\begin{equation} \label{umu}
u^{\mu} = (\cosh \eta\, \cosh \eta_r, \, \sinh \eta_r \, {\bf e}_r,\,
\sinh \eta\,  \cosh \eta_r) \,\,.
\end{equation}
Note that longitudinal boost invariance implies that $\eta_r$ cannot
depend on $\eta$, and is therefore given by the solution of the
hydrodynamical equations at $z=0$, $\eta_r \equiv {\rm Artanh}\, 
[v(z=0)]$, as provided in Section 3.

For the single inclusive momentum distribution (in Boltzmann approximation)
we insert (\ref{dsigmadotp}) and (\ref{umu}) into (\ref{single}),
perform the $\eta$--integration with the help of eq.\ (3.547.4) of \cite{GR}, 
and the $\phi$--integration using the formula (3.937.2) \cite{GR}, resulting 
in
\begin{eqnarray}
E\, \frac{{\rm d}N}{{\rm d}^3 {\bf p}} & = & \frac{d}{2 \pi^2} \int_0^1
{\rm d}\zeta\, r_f(\zeta)\, \tau_f(\zeta) \left\{ -m_\perp \,K_1 
\left(\frac{m_\perp \cosh \eta_r}{T} \right)\, 
I_0 \left(\frac{p_\perp \sinh \eta_r}{T} \right)\,
\frac{{\rm d}r_f}{{\rm d}\zeta} \right. \nonumber \\
&  & \hspace*{4.2cm} \left.
+p_\perp \,K_0 \left(\frac{m_\perp \cosh \eta_r}{T} \right)\, 
I_1 \left(\frac{p_\perp \sinh \eta_r}{T} \right) \,
\frac{{\rm d}\tau_f}{{\rm d}\zeta} \right\} \, \,  . \label{single3}
\end{eqnarray}
Note that the final spectrum respects the symmetries of the problem,
i.e., it is azimuthally symmetric and boost invariant along $z$.
For further use, let us define
\begin{equation} \label{i0b}
{\cal I}_0 (p_\perp) = \frac{(2 \pi)^3}{d}\, E\, \frac{{\rm d}N}{{\rm d}^3
{\bf p}}\,\, .
\end{equation}
For the side--correlation function for particles at $y=0$
we may choose ${\bf K}_\perp = (K,0,0)$ such that
\begin{eqnarray}
p_1^{\mu} & = & ( E,K,q_{\rm \,side}/2,0)\,\, , \\
p_2^{\mu} & = & ( E,K,-q_{\rm \,side}/2,0)\,\, , \\
K^{\mu} & = & ( E,K,0,0)\,\, , \\
q^{\mu} & = & (0,0,q_{\rm \,side},0)\,\, , 
\end{eqnarray}
where $E=[K^2 + q_{\rm \,side}^2/4 + m^2]^{1/2}$. As in the previous
case of the fireball geometry (Appendix A), one has to calculate
${\cal I}_1$ and ${\cal I}_2$. While the latter vanishes again by
symmetry, the former reads after using eq.\ (3.547.4) of \cite{GR} in the 
$\eta$--integration
\begin{eqnarray} 
{\cal I}_1 & = & 4 \pi \int_0^1 {\rm d}\zeta \, r_f(\zeta) \, \tau_f(\zeta)
\left\{ - E \,K_1 \left(\frac{E \cosh \eta_r}{T} \right)\, 
\hat{I}_0 (a,b)\, \frac{{\rm d}r_f}{{\rm d}\zeta} \right.\nonumber  \\
&  & \hspace*{3.8cm} \left.
+K \, K_0 \left(\frac{E \cosh \eta_r}{T} \right)\, 
\hat{I}_1 (a,b)\, \frac{{\rm d}\tau_f}{{\rm d}\zeta} \right\} \, \, , 
\label{i1b}
\end{eqnarray}
where $a \equiv K\, \sinh \eta_r /T$, $ b \equiv q_{\rm \,side}\, r_f$, and
the functions $\hat{I}_0,\, \hat{I}_1$ (which have to be evaluated
numerically) are defined by
\begin{eqnarray}
\hat{I}_0 (a,b) & \equiv & \frac{1}{\pi} \int_0^\pi {\rm d} \phi\,
\cosh\, [a\, \cos \phi]\, \cos\, [b\, \sin \phi] \,\, , \\
\hat{I}_1 (a,b) & \equiv & \frac{1}{\pi} \int_0^\pi {\rm d} \phi\,
\cos \phi\, \sinh\, [a\, \cos \phi]\, \cos\, [b\, \sin \phi] \equiv 
\frac{\partial \hat{I}_0(a,b)}{\partial a}\,\, .
\end{eqnarray}
It is obvious (see eqs.\ (8.411.1, 8.431.4) in \cite{GR}) that $\hat{I}_0$
is related to the Bessel functions $I_0, \, J_0$:
\begin{equation}
\hat{I}_0 (a,0) = I_0(a)\,\,, \,\,\,\, \hat{I}_0(0,b) = J_0 (b)\,\,.
\end{equation}
The final result for the side--correlation function reads (as in the
preceding Appendix)
\begin{equation}
C_{2,{\rm \,side}} = 1 + {\cal I}_1^2/{\cal I}_0^2\,\, ,
\end{equation}
with ${\cal I}_1$ from (\ref{i1b}) and ${\cal I}_0$ from
(\ref{i0b}) with the single inclusive spectrum (\ref{single3}).

For the out--correlation the choice of momenta is
\begin{eqnarray}
p_1^{\mu} & = & ( E_1,K+q_{\rm \, out}/2,0,0)\,\, , \\
p_2^{\mu} & = & ( E_2,K-q_{\rm \, out}/2,0,0)\,\, , \\
K^{\mu} & = & ( K^0,K,0,0)\,\, , \\
q^{\mu} & = & (E_1-E_2,q_{\rm \,out},0,0)\,\, ,
\end{eqnarray}
where $E_{1,2}=[(K \pm q_{\rm \,out}/2)^2 + m^2]^{1/2},\,K^0 = (E_1+E_2)/2$. 
As in the previous cases, the $\eta$-- and $\phi$--integrations 
separate, and the final result for ${\cal I}_1$ and ${\cal I}_2$ reads
\begin{eqnarray}
{\cal I}_1 & = & 4 \pi \int_0^1 {\rm d}\zeta \, r_f(\zeta) \, \tau_f(\zeta)
\left\{ - K^0\, \left[ {\cal K}_1(\alpha,\beta)\, 
{\cal J}_0 (a,b) + \hat{\cal K}_1(\alpha,\beta)\,
\hat{\cal J}_0(a,b) \right] \, \frac{{\rm d}r_f}{{\rm d}\zeta} \right. 
\nonumber \\
&  & \hspace*{3.9cm} \left.
+K \,\left[{\cal K}_0 (\alpha,\beta)\,
{\cal J}_1 (a,b) + \hat{\cal K}_0(\alpha,\beta)\,
\hat{\cal J}_1(a,b) \right]\, \frac{{\rm d}\tau_f}{{\rm d}\zeta} \right\} 
\, \, , \label{i1c} \\
{\cal I}_2 & = & 4 \pi \int_0^1 {\rm d}\zeta \, r_f(\zeta) \, \tau_f(\zeta)
\left\{ - K^0\, \left[ \hat{\cal K}_1(\alpha,\beta)\, 
{\cal J}_0 (a,b) - {\cal K}_1(\alpha,\beta)\,
\hat{\cal J}_0(a,b) \right] \, \frac{{\rm d}r_f}{{\rm d}\zeta} \right. 
\nonumber \\
&  & \hspace*{3.9cm} \left.
+K \,\left[\hat{\cal K}_0 (\alpha,\beta)\,
{\cal J}_1 (a,b) - {\cal K}_0(\alpha,\beta)\,
\hat{\cal J}_1(a,b) \right]\, \frac{{\rm d}\tau_f}{{\rm d}\zeta} \right\} 
\, \, , \label{i2c}
\end{eqnarray}
where $\alpha \equiv K^0 \cosh \eta_r/T$, $\beta \equiv (E_1-E_2) \tau_f$, 
$a \equiv K\, \sinh \eta_r/T$, $b \equiv q_{\rm \,out}\, r_f$, and
(see eqs.\ (8.431.4, 8.431.5, 8.432.1, 8.476.4) in \cite{GR})
\begin{eqnarray}
{\cal K}_0 (\alpha,\beta) & \equiv & \int_0^\infty {\rm d} \eta\, 
\cos\, [\beta \, \cosh \eta]\, e^{-\alpha\, \cosh \eta} 
\equiv {\rm Re}\, K_0(\alpha-i\beta) \,\, , \\
{\cal K}_1 (\alpha,\beta) & \equiv & \int_0^\infty {\rm d} \eta\, \cosh \eta\, 
\cos\, [\beta \, \cosh \eta]\, e^{-\alpha\, \cosh \eta} 
\equiv {\rm Re}\, K_1(\alpha-i\beta) \equiv 
- \frac{\partial {\cal K}_0 (\alpha,\beta)}{\partial \alpha}\,\, , \\
\hat{\cal K}_0 (\alpha,\beta) & \equiv & \int_0^\infty {\rm d} \eta\, 
\sin\, [\beta \, \cosh \eta]\, e^{-\alpha\, \cosh \eta} 
\equiv {\rm Im}\, K_0(\alpha-i\beta) \,\, , \\
\hat{\cal K}_1 (\alpha,\beta) & \equiv & \int_0^\infty {\rm d} \eta\, 
\cosh \eta\, 
\sin\, [\beta \, \cosh \eta]\, e^{-\alpha\, \cosh \eta} 
\equiv {\rm Im}\, K_1(\alpha-i\beta) \equiv 
- \frac{\partial \hat{\cal K}_0 (\alpha,\beta)}{\partial \alpha}\,\,, \\
{\cal J}_0 (a,b) & \equiv & \frac{1}{\pi} \int_0^\pi {\rm d} \phi\, 
\cos\, [b \cos \phi]\, \cosh\, [a\, \cos \phi]
\equiv {\rm Re}\, I_0(a+ib) \,\, , \\
{\cal J}_1 (a,b) & \equiv & \frac{1}{\pi} \int_0^\pi {\rm d} \phi\, 
\cos \phi\, \cos\, [b \cos \phi]\, \sinh\, [a\, \cos \phi] 
\equiv {\rm Re}\, I_1(a+ib) \equiv
\frac{\partial {\cal J}_0(a,b)}{\partial a} \,\, , \\
\hat{\cal J}_0 (a,b) & \equiv & \frac{1}{\pi} \int_0^\pi {\rm d} \phi\, 
\sin\, [b \cos \phi]\, \sinh\, [a\, \cos \phi]
\equiv {\rm Im}\, I_0(a+ib) \,\, , \\
\hat{\cal J}_1 (a,b) & \equiv & \frac{1}{\pi} \int_0^\pi {\rm d} \phi\, 
\cos \phi\, \sin\, [b \cos \phi]\, \cosh\, [a\, \cos \phi]
\equiv {\rm Im}\, I_1(a+ib) \equiv
\frac{\partial \hat{\cal J}_0(a,b)}{\partial a} \,\, .
\end{eqnarray}
The final result for the out--correlation reads
\begin{equation}
C_{2,{\rm \,out}} = 1 + \frac{{\cal I}_1^2 + {\cal I}_2^2}{{\cal I}_0
(p_{1,\perp})\,\, {\cal I}_0(p_{2,\perp})} \,\, ,
\end{equation}
as before, cf.\ (\ref{c2out}), but with ${\cal I}_1,\, {\cal I}_2$ 
from (\ref{i1c}) and (\ref{i2c}) and ${\cal I}_0$ from (\ref{i0b}) with
the single inclusive spectrum (\ref{single3}).

\newpage
\noindent 
{\bf Figure Captions:}
\\ ~~ \\ 
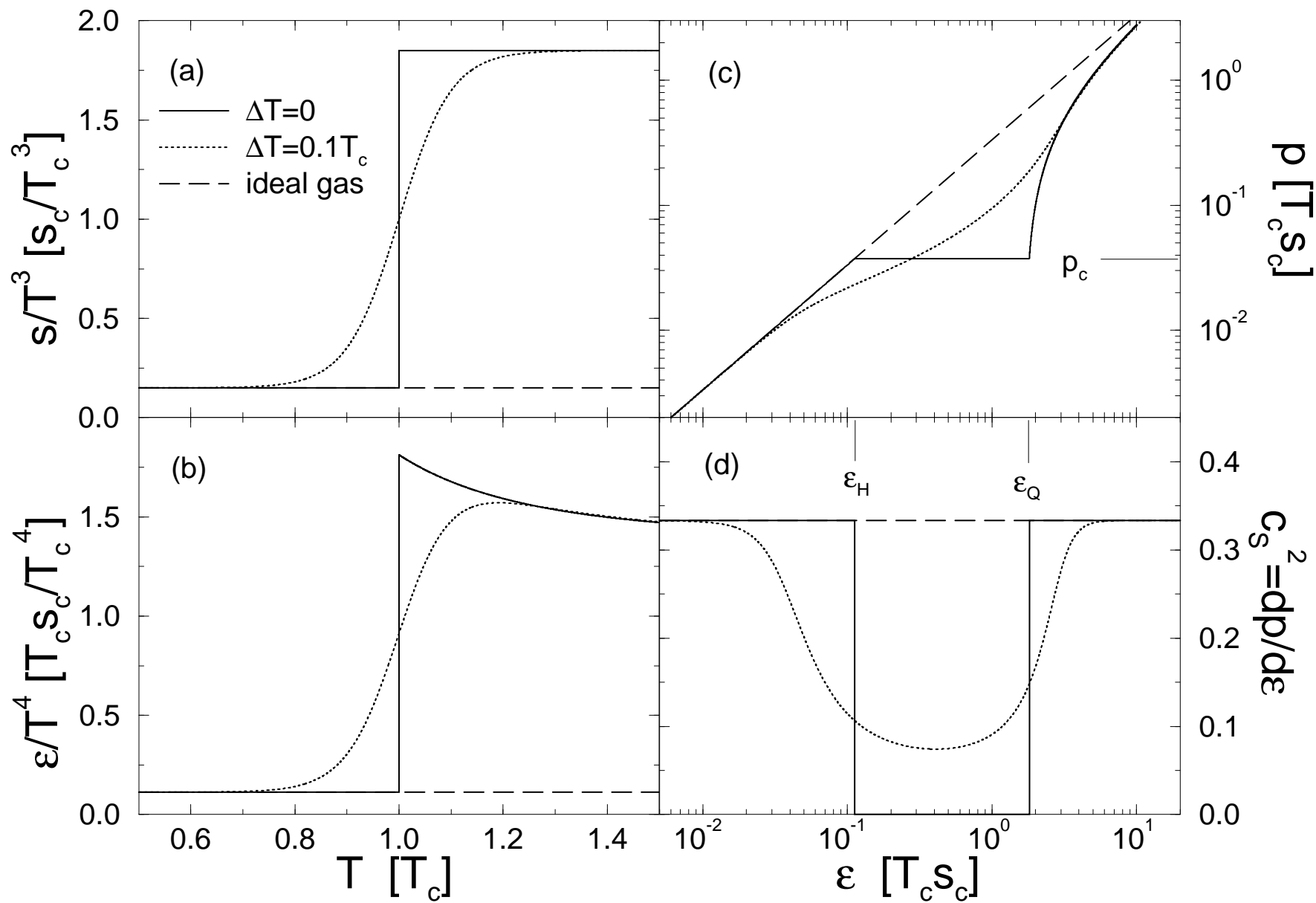
{\bf Fig.\ 1:} (a) the entropy density divided by $T^3$
(in units of $s_c/T_c^3$), (b) the energy density divided by 
$T^4$ (in units of $T_c\, s_c/ T_c^4$) as functions of temperature
(in units of $T_c$),
(c) the pressure (in units of $T_c\, s_c$), (d) the square of
the velocity of sound as functions of energy density
(in units of $T_c\, s_c$). The solid lines
correspond to $\Delta T=0$, the dotted curves to
$\Delta T=0.1\, T_c$. Quantities for the ideal gas equation of state
(with $d_H$ degrees of freedom) are represented by dashed lines.
The ratio of degrees of freedom in the QGP to those in the hadronic phase is
$d_Q/d_H=37/3$. The critical enthalpy density
is $T_c\, s_c \simeq  0.75$ GeV fm$^{-3}$ for the case $d_Q=37,\, d_H=3$. 
\\ ~~ \\ 
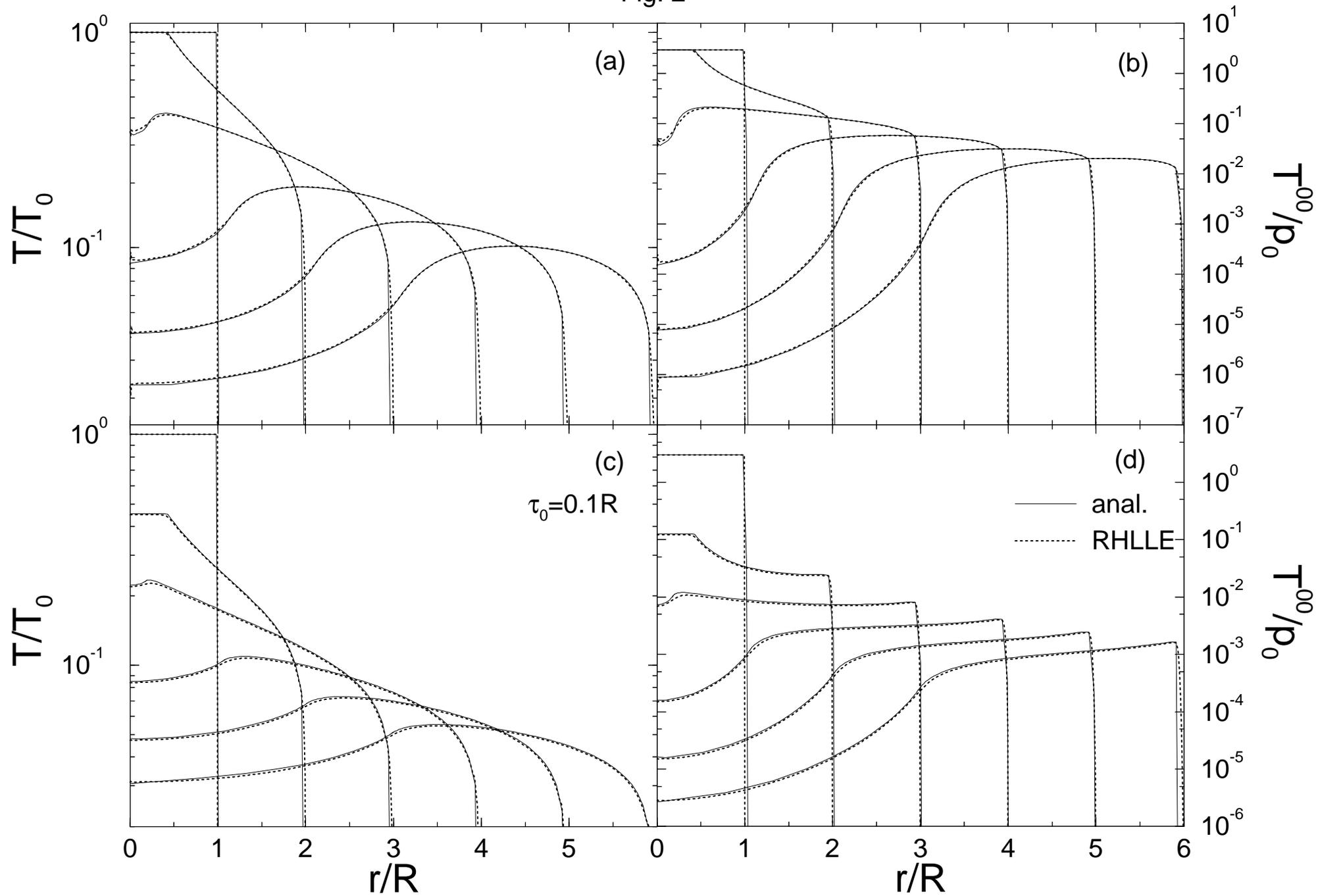
{\bf Fig.\ 2:} The expansion of (a,b) a fireball and (c,d) a Bjorken cylinder
(with initial time $\tau_0=0.1\, R$)
for an ideal gas equation of state $p=\epsilon/3$. (a,c) temperature 
(in units of the initial temperature $T_0$), (b,d) laboratory
energy density (in units of the initial pressure $p_0$) as functions
of radial distance $r$ from the origin (in units of the initial radius
$R$ of the system). The profiles are for times $t=\tau_0+n \lambda R$, 
$n=0,1,2,...,5$, $\lambda = 0.99$. Solid lines are obtained via
the semi-analytical method of characteristics \cite{baym}, dotted curves are
produced with the relativistic HLLE algorithm modified with Sod's operator
splitting method.
\\ ~~ \\ 
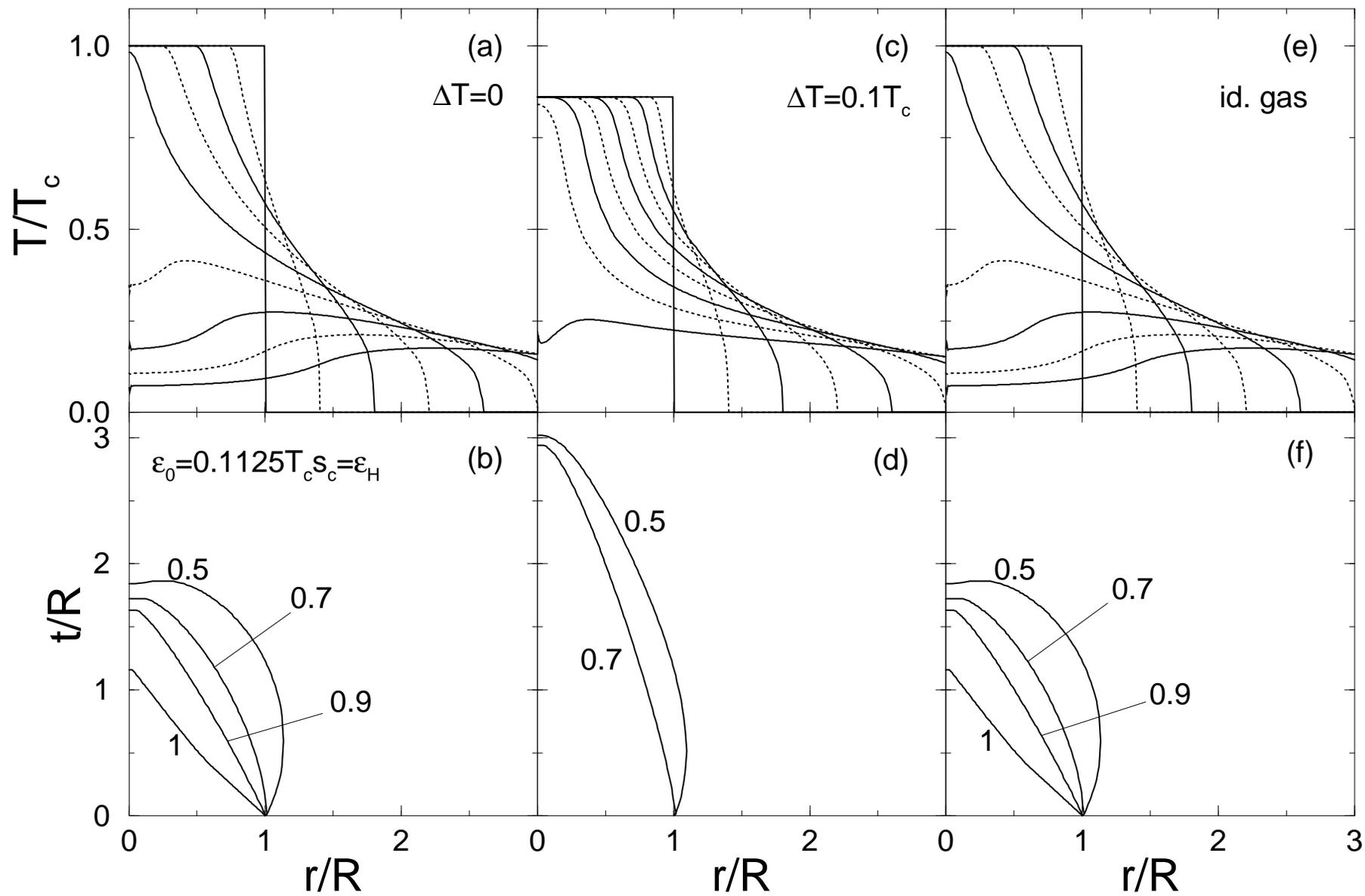
{\bf Fig.\ 3:} (a,c,e) temperature profiles for the fireball expansion
for times $t=0.4\, n \lambda R,\, n=0,1,...,8$ and an initial energy
density $\epsilon_0 = \frac{3}{2}\, T_c\, s_c/\, (d_Q/d_H+1) = 
0.1125\, T_c\, s_c \equiv \epsilon_H$. The profiles
are alternatingly shown as full and dotted lines in order to better
distinguish them. (b,d,f) show isotherms in the corresponding 
space--time diagrams. The isotherms are labelled with the
corresponding temperatures in units of $T_c$. Figs.\ 
(a,b) are calculated with $\Delta T=0$, (c,d) with
$\Delta T=0.1\, T_c$, and (e,f) with the ideal gas equation of state.
\\ ~~ \\ 
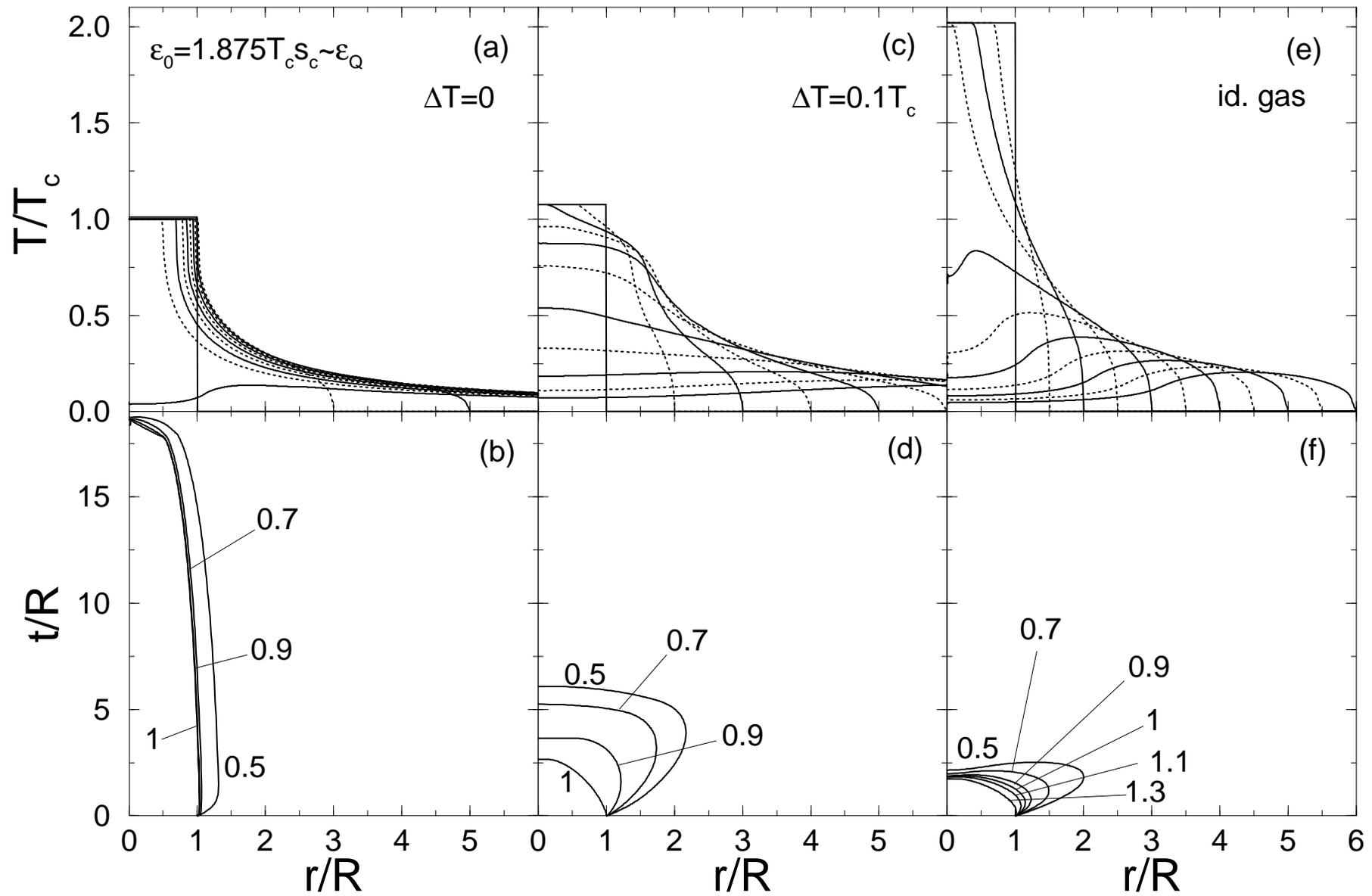
{\bf Fig.\ 4:} Same as in Fig.\ 3, for 
$\epsilon_0 = 1.875\, T_c\, s_c \sim \epsilon_Q$.
Profiles in (a) are for times $t=2\, n \lambda R$,
in (c) for $t=n \lambda R$, and in (e)
for $t=0.5\, n \lambda R,\, n=0,1,...,10$.
\\ ~~ \\ 
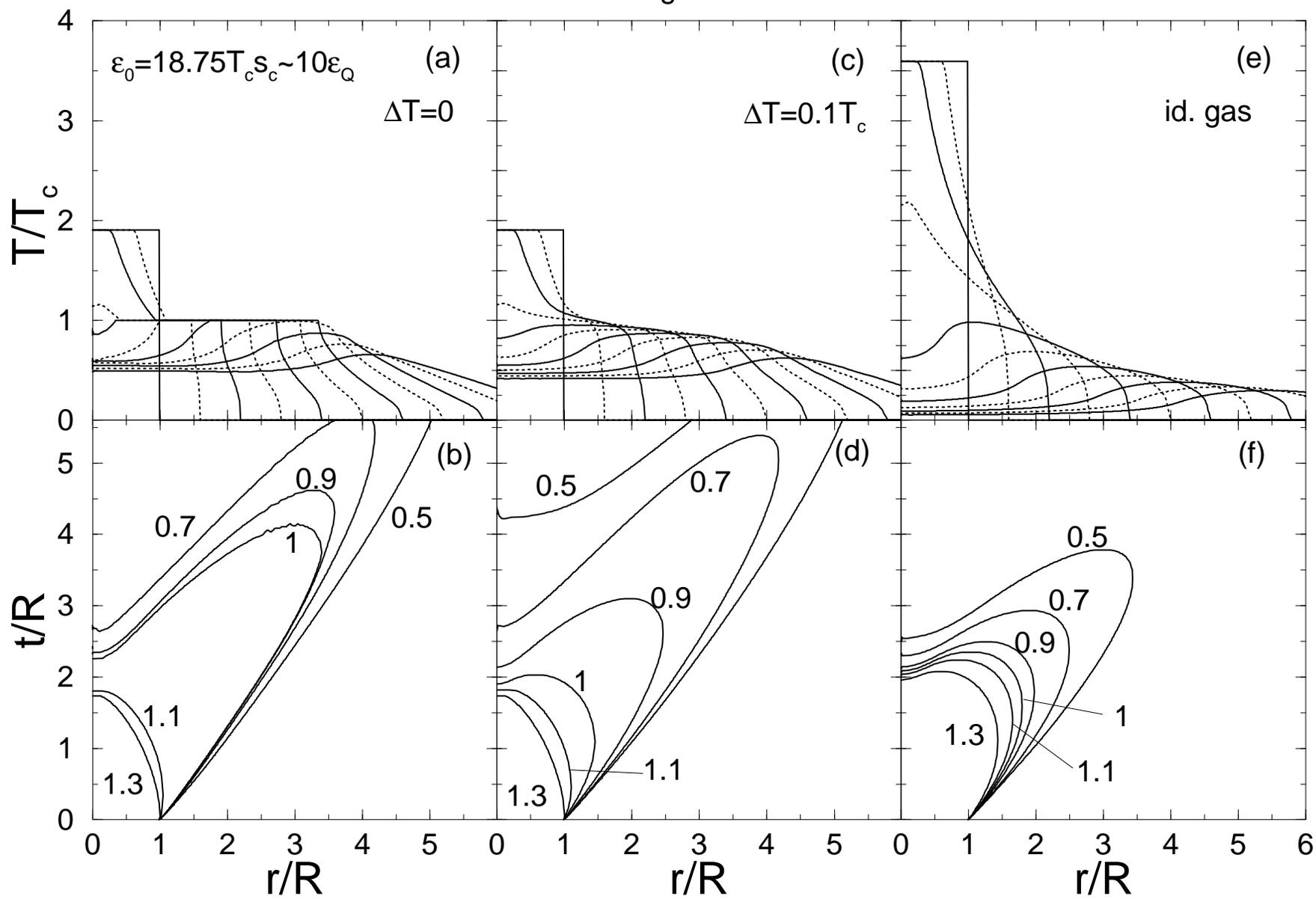
{\bf Fig.\ 5:} Same as in Fig.\ 3, for 
$\epsilon_0 = 18.75\, T_c\, s_c \sim 10\, \epsilon_Q$. 
Profiles in (a,c,e) are for times $t=0.6\, n \lambda
R,\, n=0,1,...,10$.
\\ ~~ \\ 
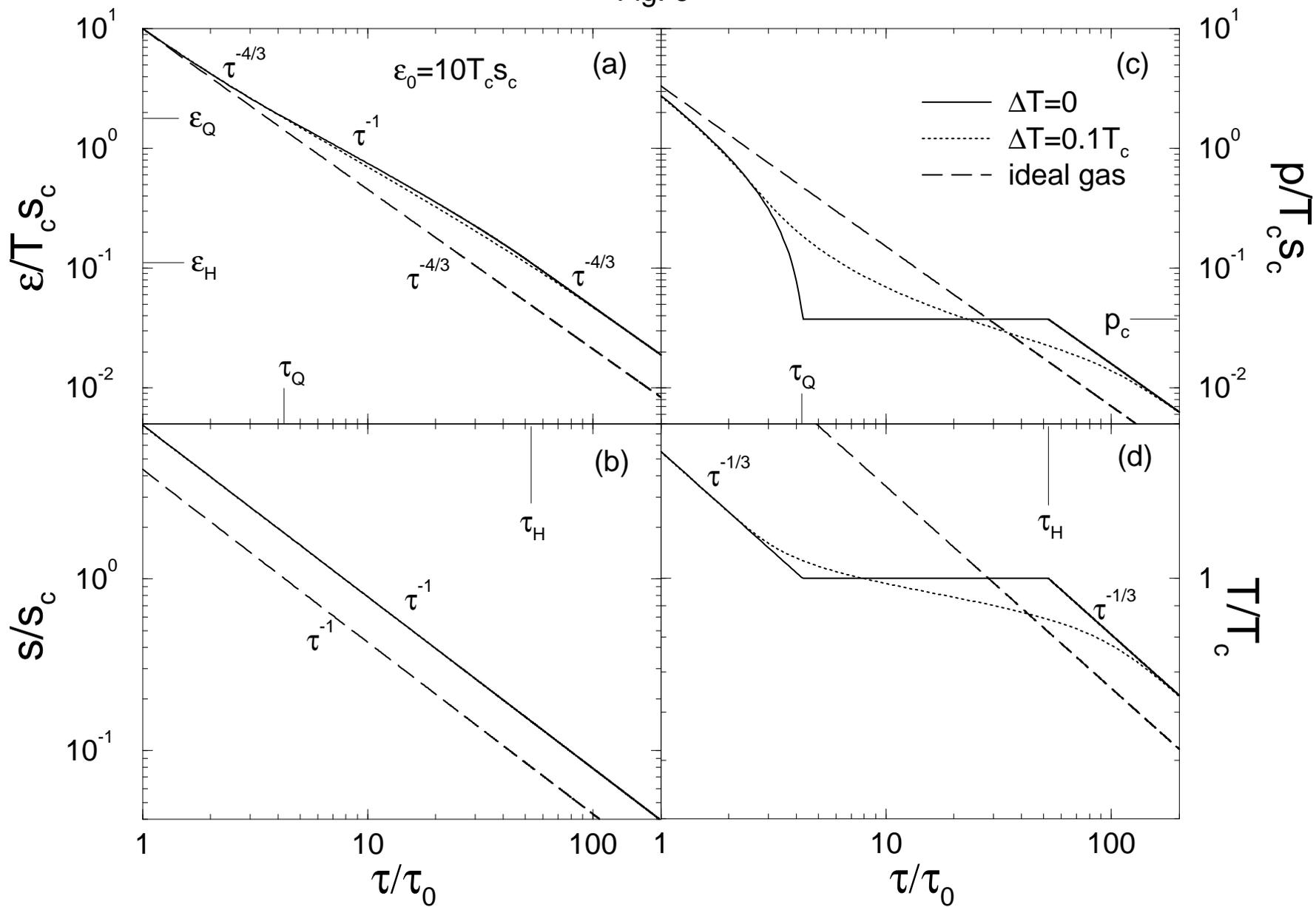
{\bf Fig.\ 6:} Purely longitudinal expansion
in the Bjorken model for an initial energy density
$\epsilon_0 = 10\, T_c\, s_c$. (a) energy density (in units of $T_c\, s_c$),
(b) entropy density (in units of $s_c$), (c) pressure (in units
of $T_c\, s_c$), (d) temperature (in units of $T_c$)
as functions of (proper) time $\tau$ (in units of $\tau_0$).
Solid lines are for $\Delta T=0$, dotted for $\Delta T=0.1\, T_c$,
dashed for the ideal gas.
Different cooling laws $\tau^\alpha$ are indicated.
\\ ~~ \\ 
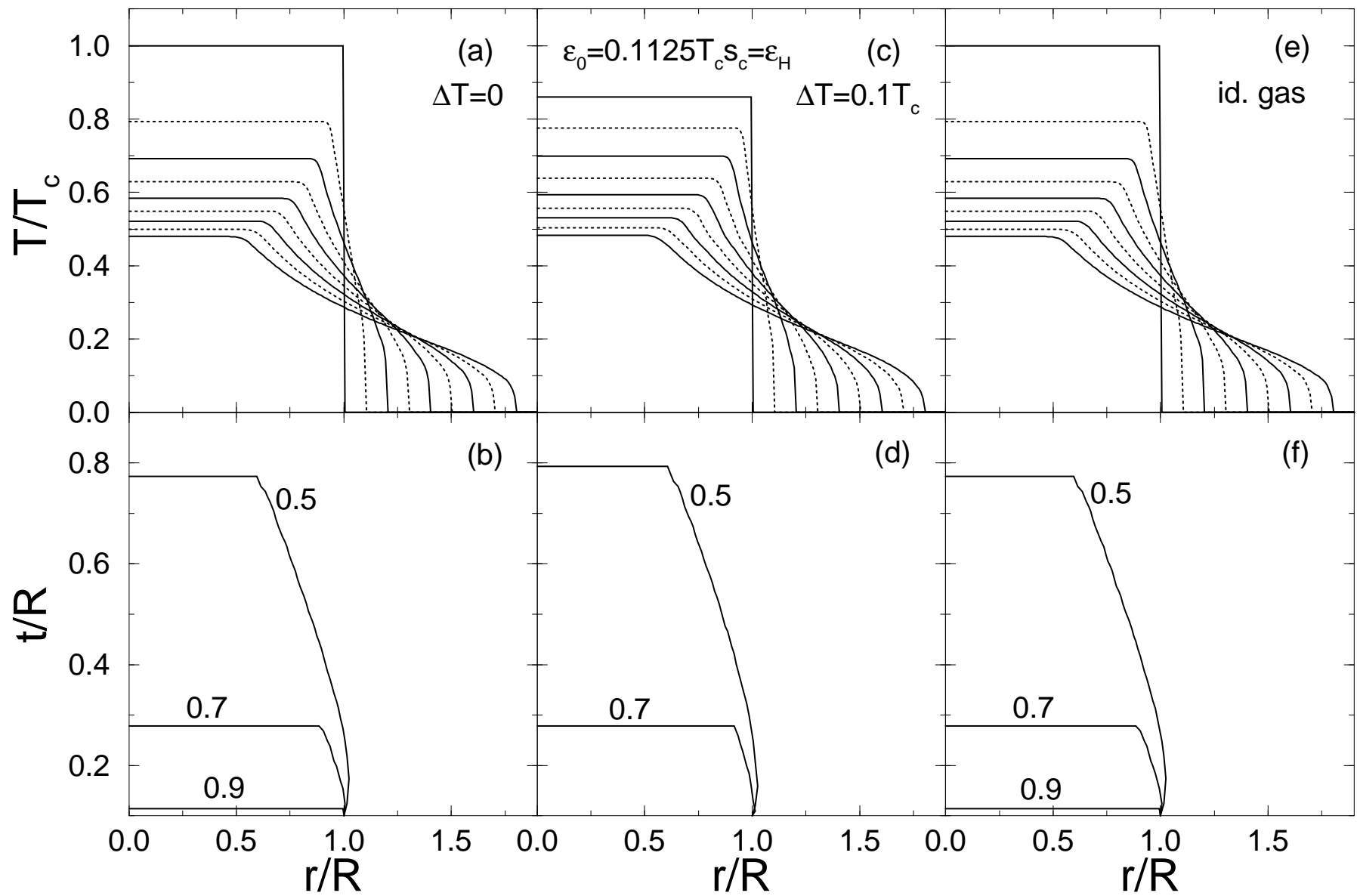
{\bf Fig.\ 7:} Same as in Fig.\ 3, but for the Bjorken cylinder
expansion with $\tau_0=0.1\, R$. Profiles in (a,c,e) are
for times $t=\tau_0+0.1\, n \lambda R,\, n=0,1,...,8$. 
\\ ~~ \\ 
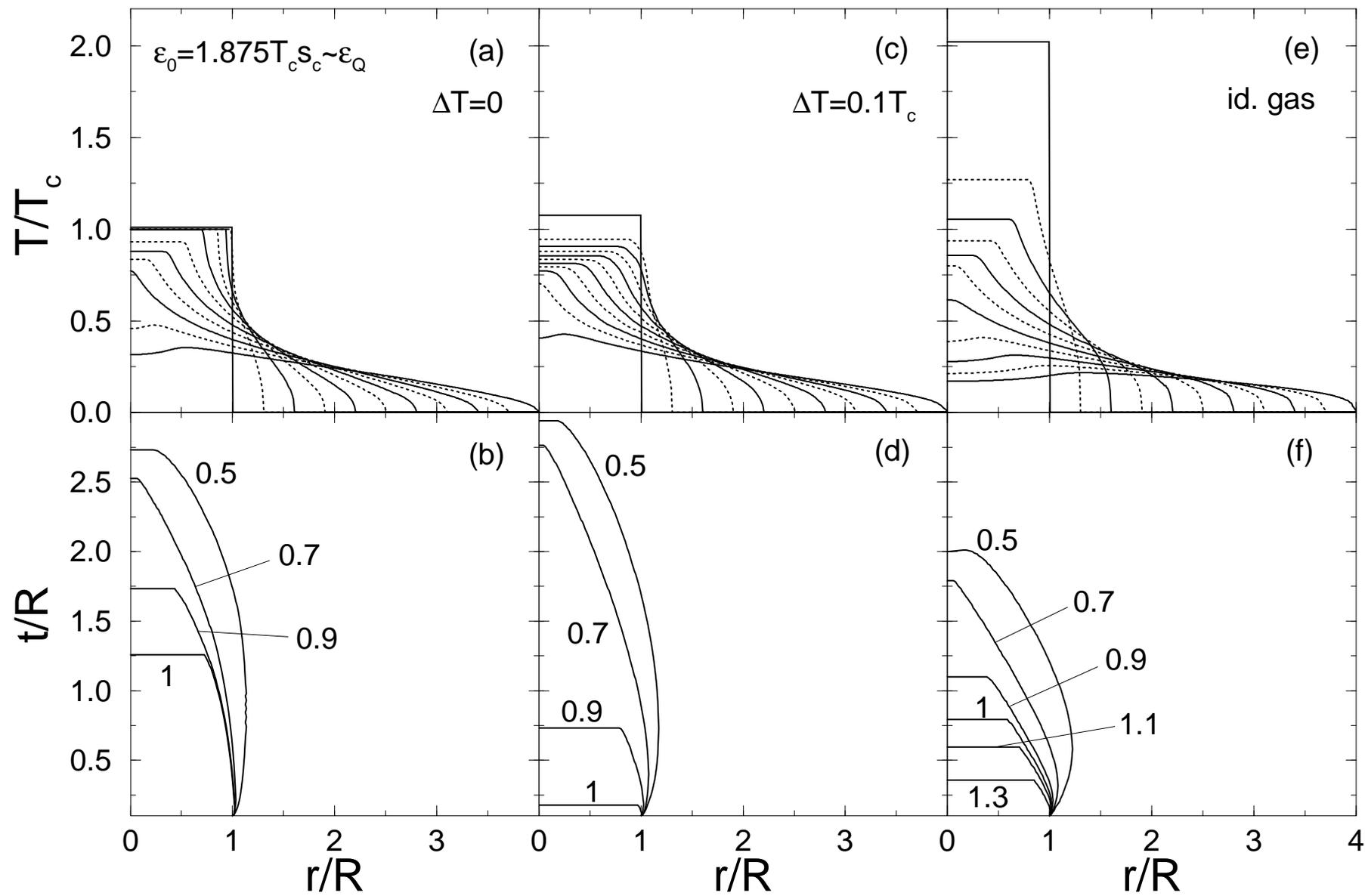
{\bf Fig.\ 8:} Same as in Fig.\ 7, for an initial energy density
$\epsilon_0 = 1.875\, T_c\, s_c \sim \epsilon_Q$.
Profiles in (a,c,e) are for times $t=\tau_0 + 0.3\, n \lambda R,\,
n=0,1,...,10$.
\\ ~~ \\
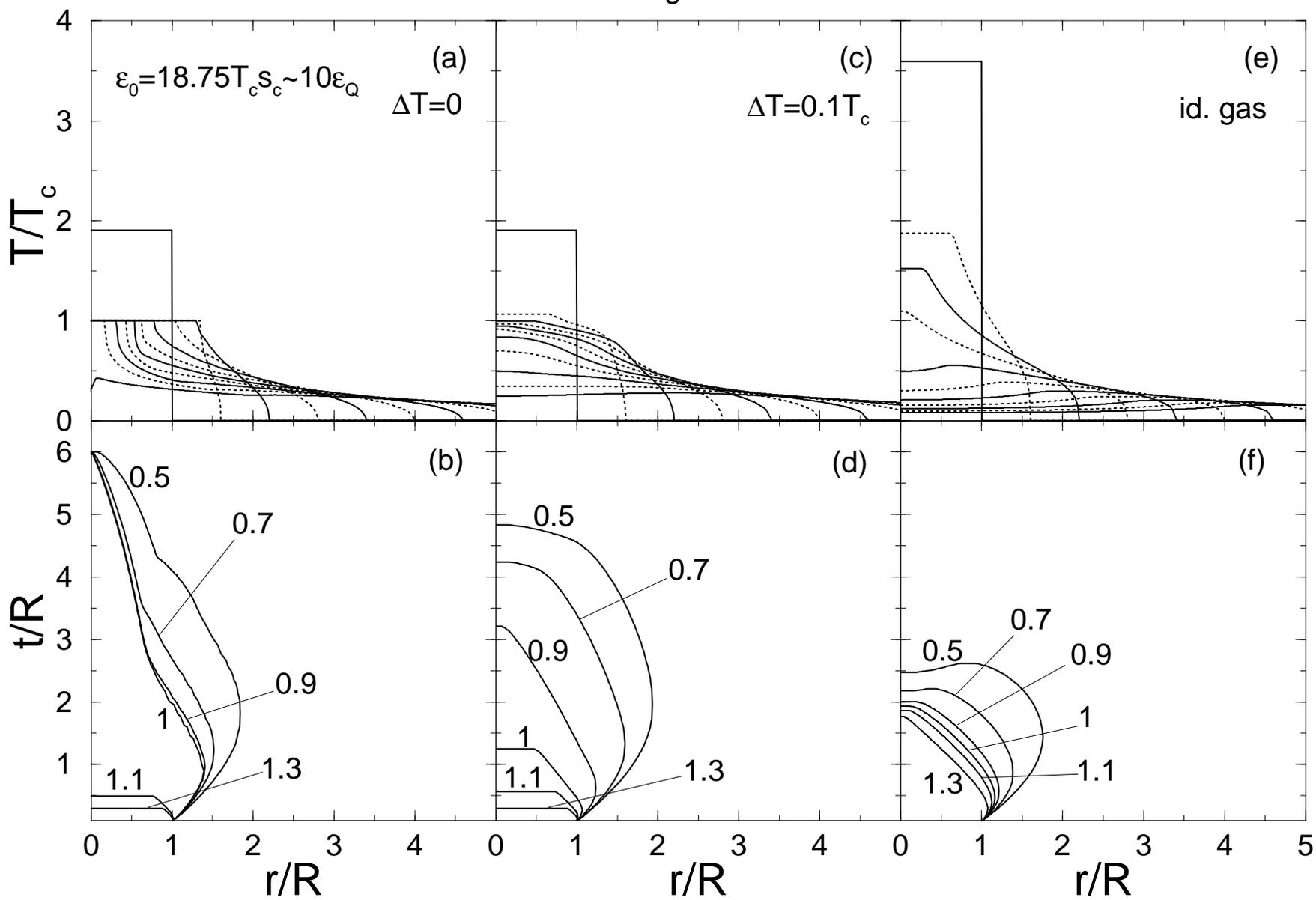
{\bf Fig.\ 9:} Same as in Fig.\ 7, for an initial energy density
$\epsilon_0 = 18.75\, T_c\, s_c \sim 10\, \epsilon_Q$.
Profiles in (a,c,e) are for times $t=\tau_0 + 0.6\, n \lambda R,\,
n=0,1,...,10$.
\\ ~~ \\
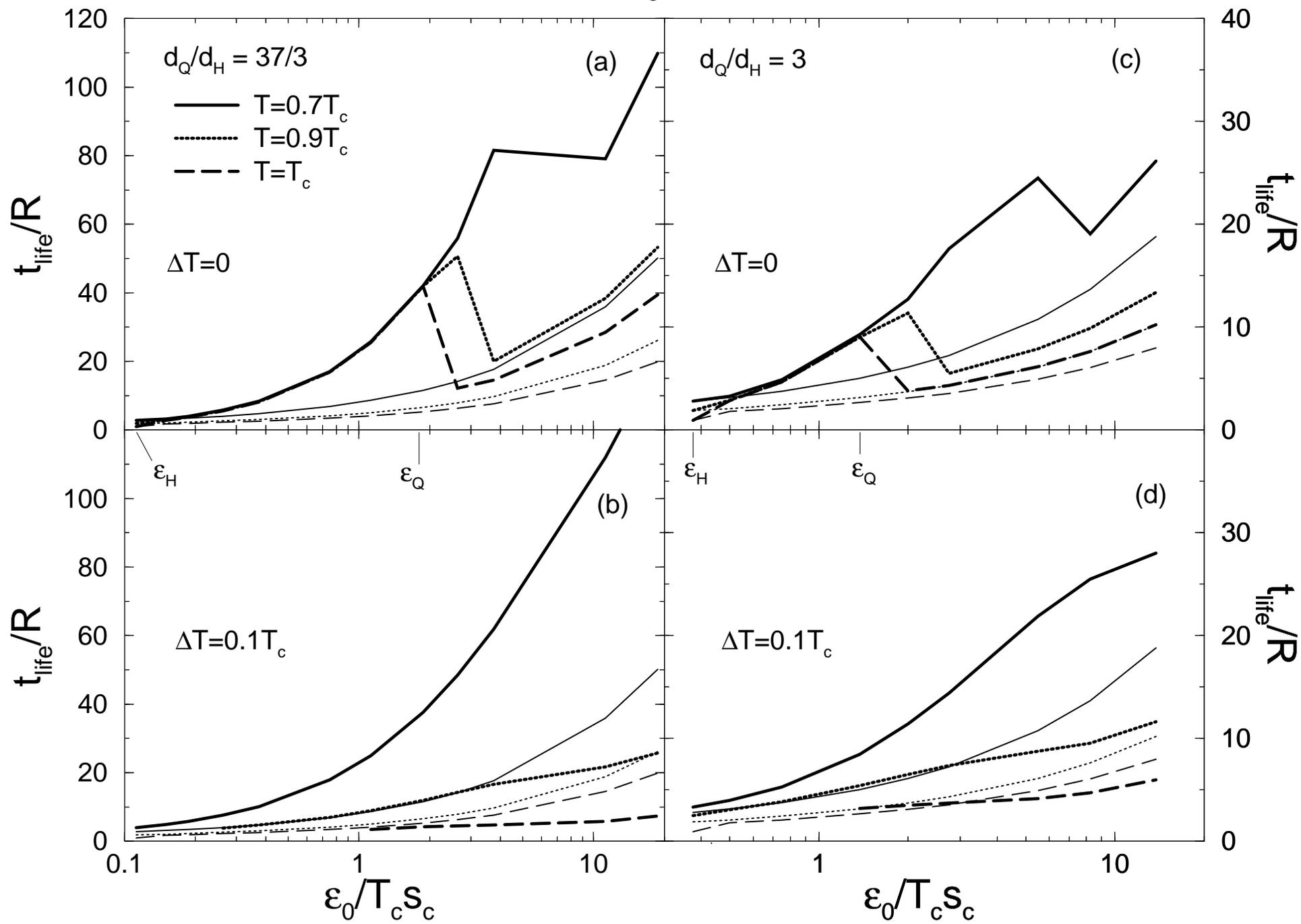
{\bf Fig.\ 10:} Lifetimes (in units of the initial radius $R$ of the system)
in the one--dimensional (Landau) expansion (cf.\ also \cite{dhrmg})
as a function of the initial energy density (in units of $T_c\, s_c$).
(a,b) are for $d_Q/d_H = 37/3$, (c,d) for $d_Q/d_H=3$.
The thick lines in (a,c) are for $\Delta T=0$, in (b,d) 
for $\Delta T=0.1\, T_c$. Thin lines correspond to the ideal gas case.
Solid lines are for $T=0.7\, T_c$, dotted for $\Delta T=0.9\, T_c$,
dashed for $T=T_c$.
\\ ~~ \\
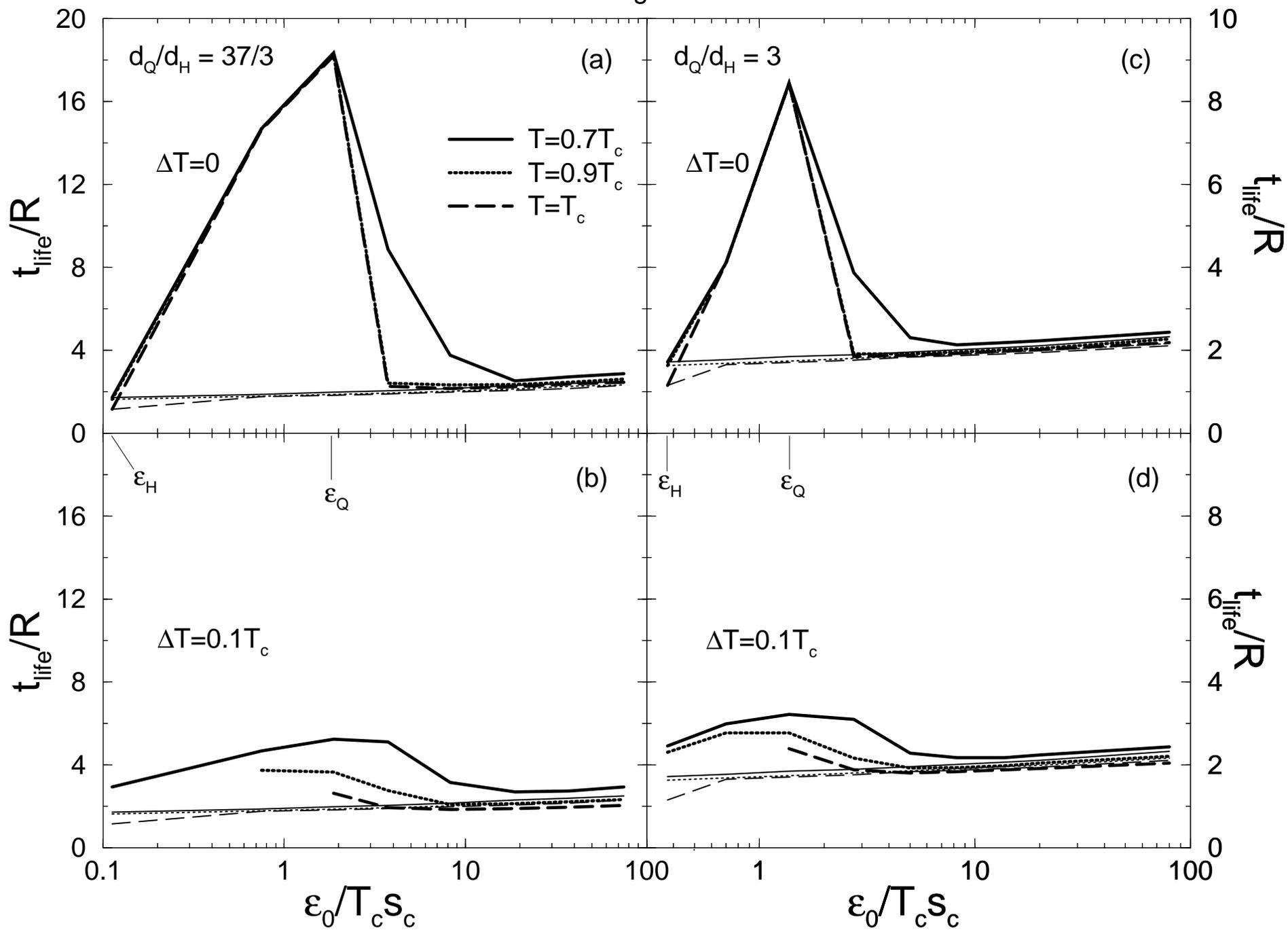
{\bf Fig.\ 11:} The same as in Fig.\ 10, but for the fireball expansion.
\\ ~~ \\
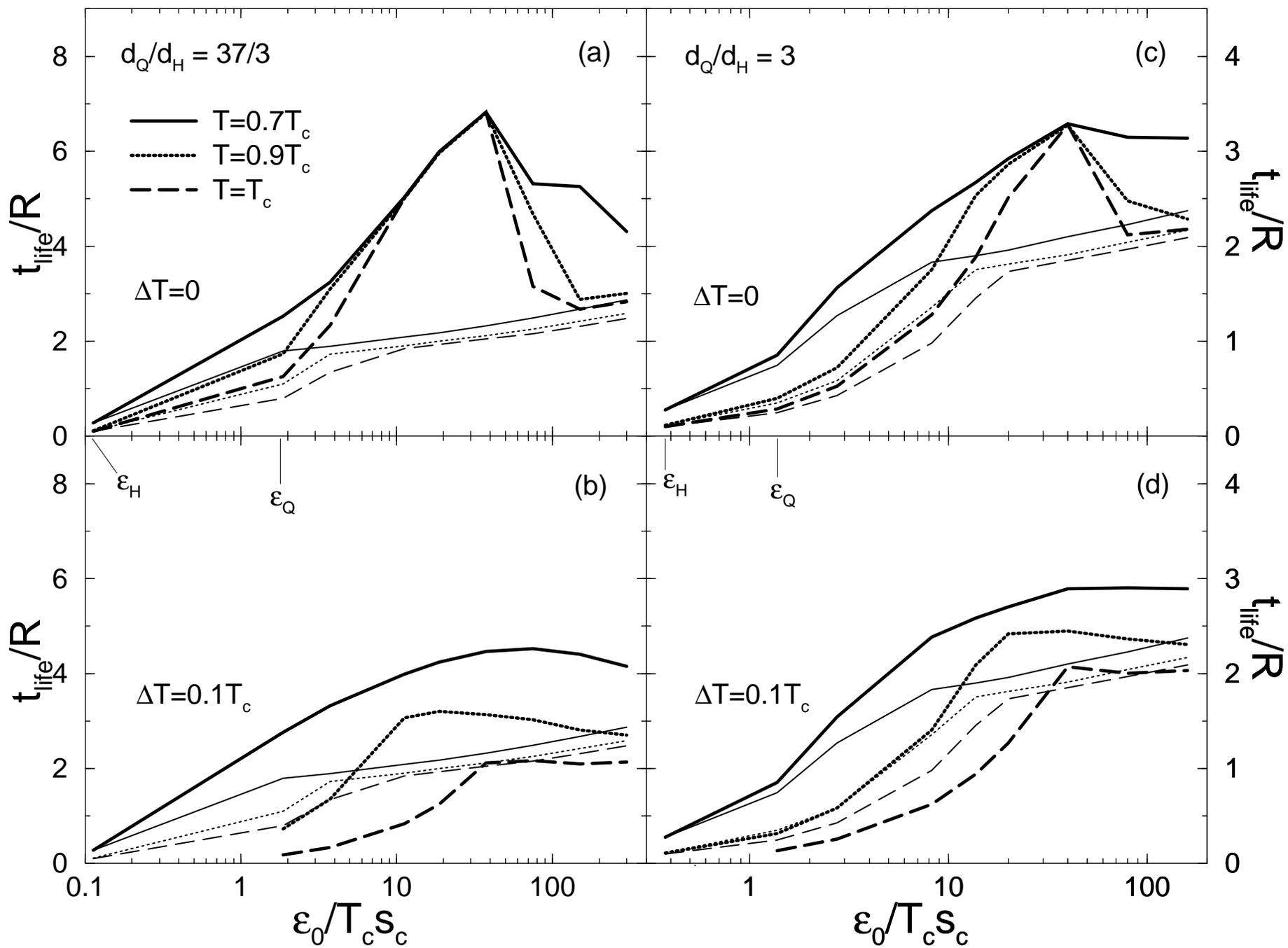
{\bf Fig.\ 12:} The same as in Fig.\ 10, but for the Bjorken cylinder
expansion with $\tau_0= 0.1\, R$.
\\ ~~ \\
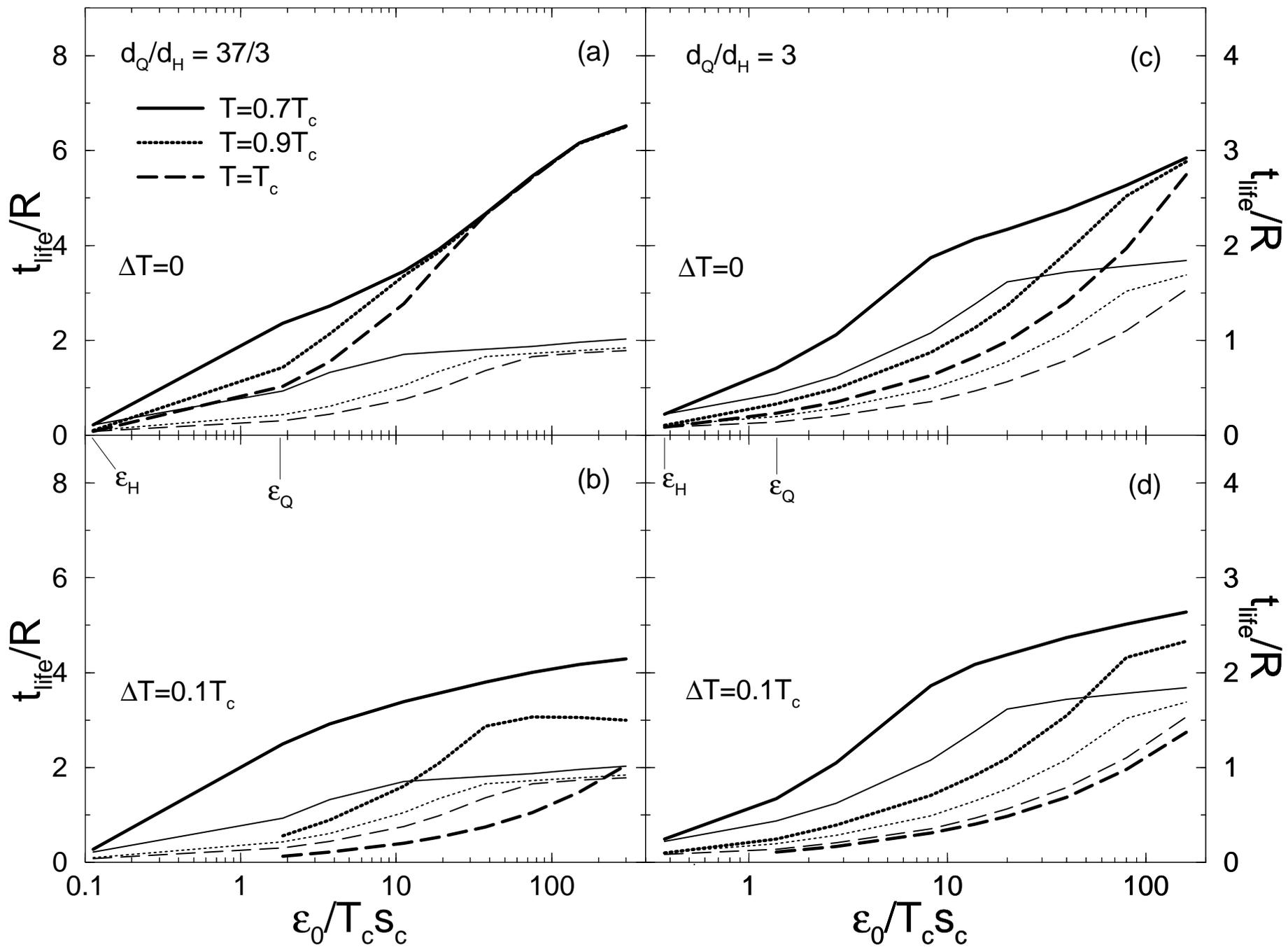
{\bf Fig.\ 13:} The same as in Fig.\ 12, but for $\tau_0= 1/3\, T_0$.
\\ ~~ \\
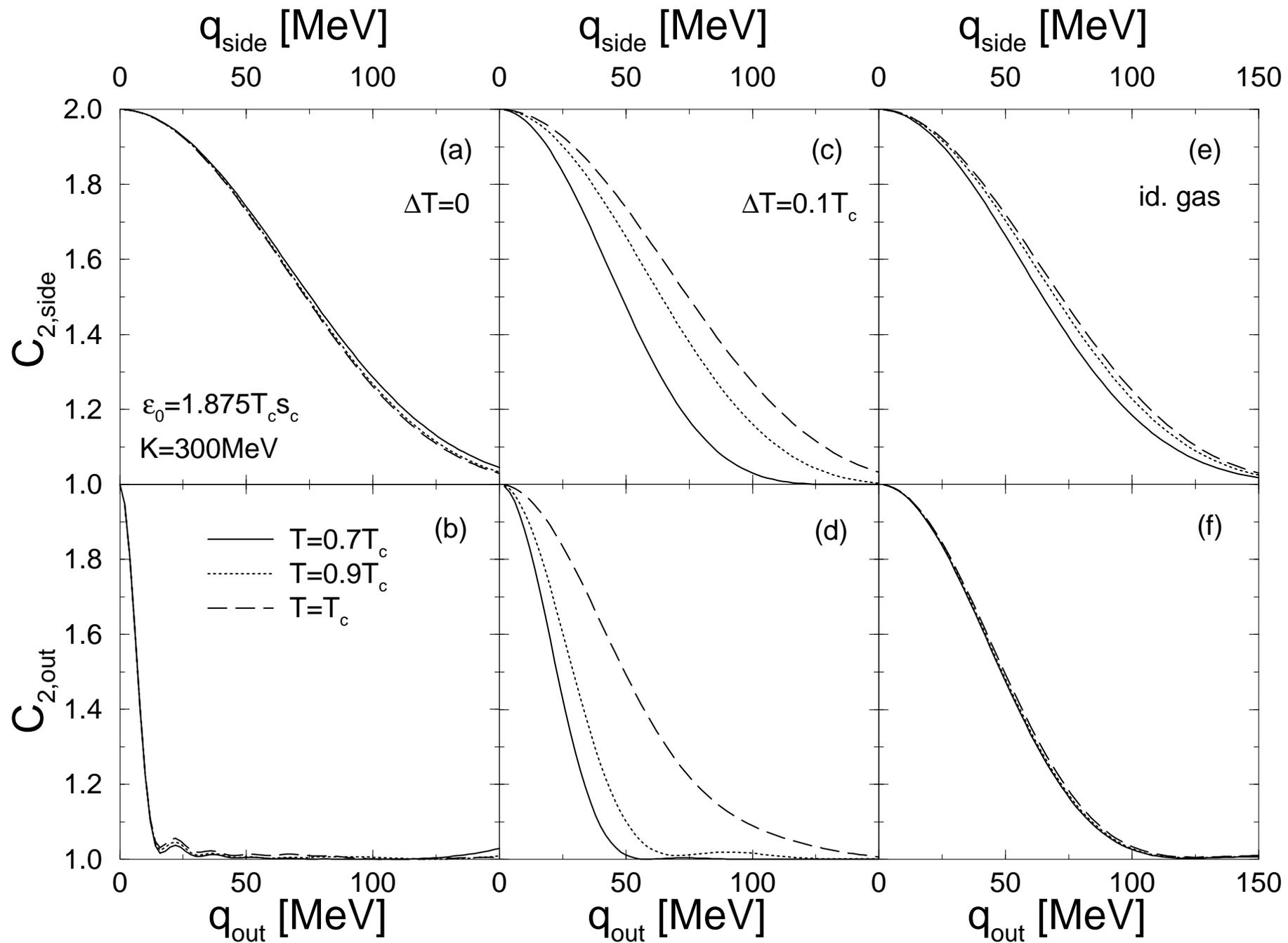
{\bf Fig.\ 14:} Two--pion correlation functions in (a,c,e) side--
and (b,d,f) out--direction for the fireball expansion with
$\epsilon_0=1.875\, T_c\, s_c$. The average pion momentum
is $K=300$ MeV, the initial radius $R$ was fixed to be 5 fm. The
critical temperature was taken as $T_c= 160$ MeV, the pion mass is
$m=138$ MeV. (a,b) are for
$\Delta T=0$, (c,d) for $\Delta T=0.1\, T_c$, (e,f) for
the ideal gas. Solid, dotted, and dashed 
lines are the correlation functions as calculated along 
the $T=0.7\, T_c,\, 0.9\, T_c$, and $T_c$ isotherm, respectively.
\\ ~~ \\
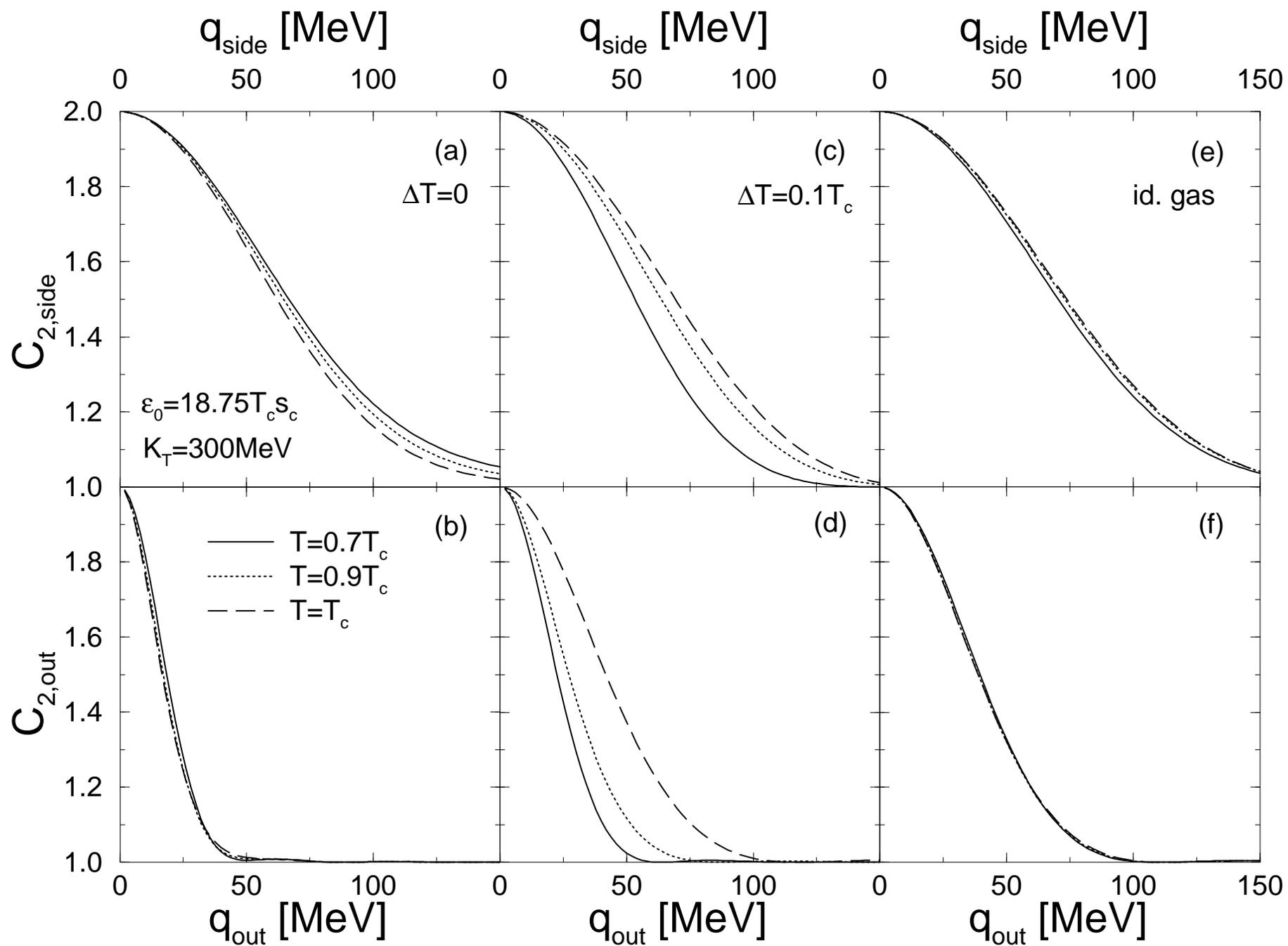
{\bf Fig.\ 15:} The same as in Fig.\ 14, but for the
Bjorken cylinder expansion with $\tau_0=0.1\, R$ and
$\epsilon_0=18.75\, T_c\, s_c$. The average transverse pion momentum
is $K=300$ MeV, the longitudinal momenta of the pions vanish.
\\ ~~ \\
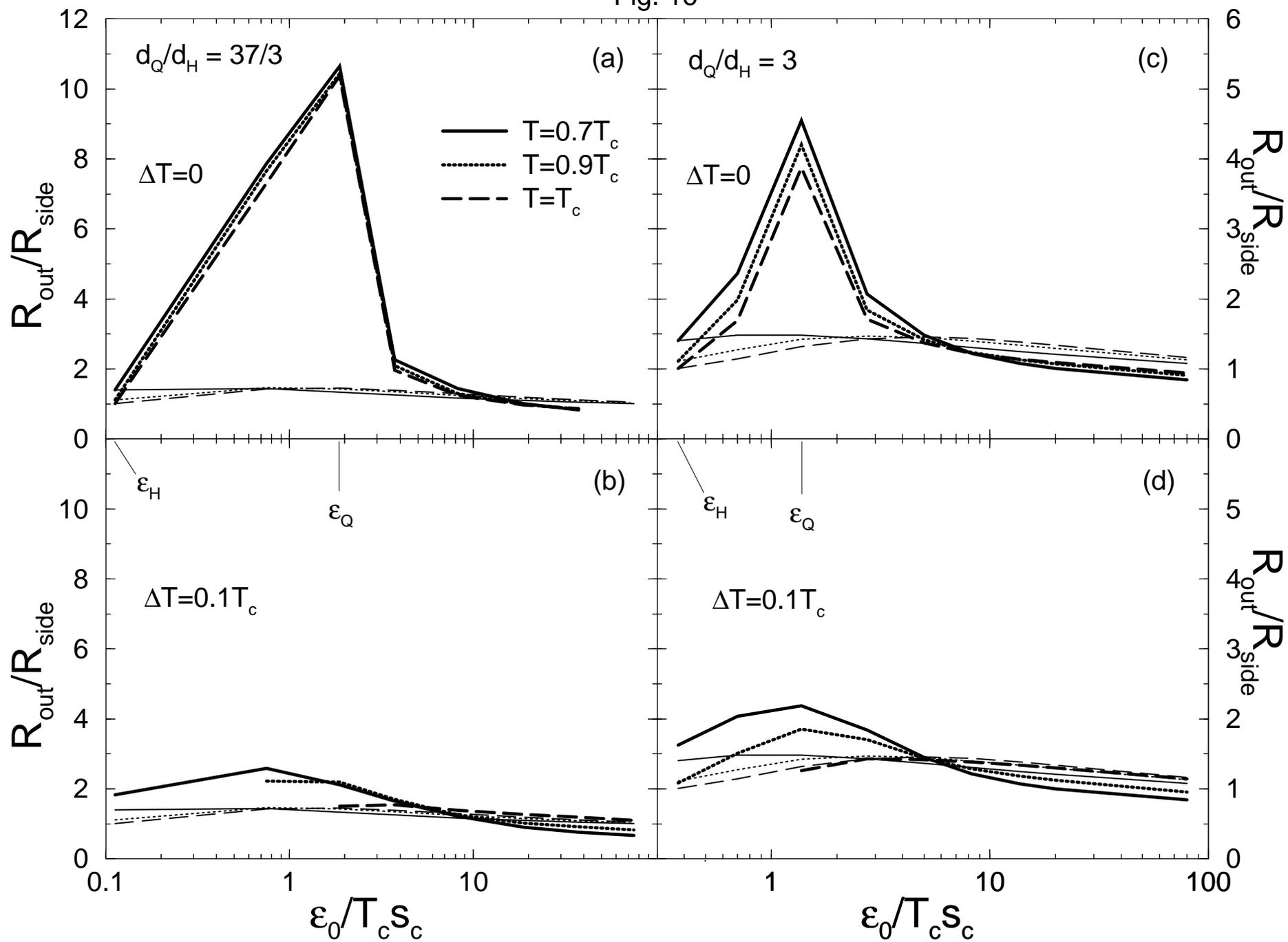
{\bf Fig.\ 16:} The same as in Fig.\ 11, but for the ratio
$R_{\rm \,out}/R_{\rm \,side}$.
\\ ~~ \\
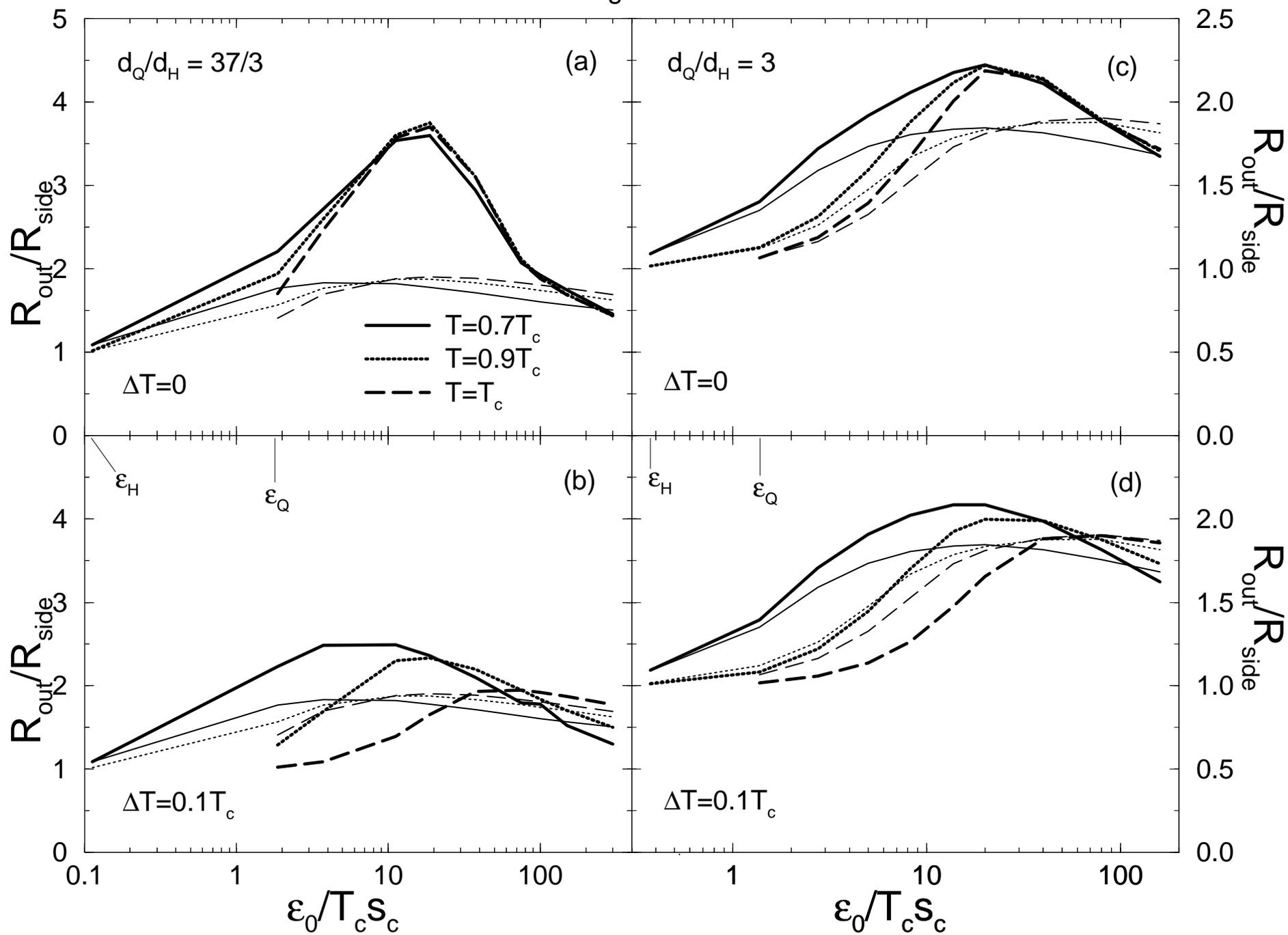
{\bf Fig.\ 17:} The same as in Fig.\ 12, but for the ratio
$R_{\rm \,out}/R_{\rm \,side}$.
\\ ~~ \\
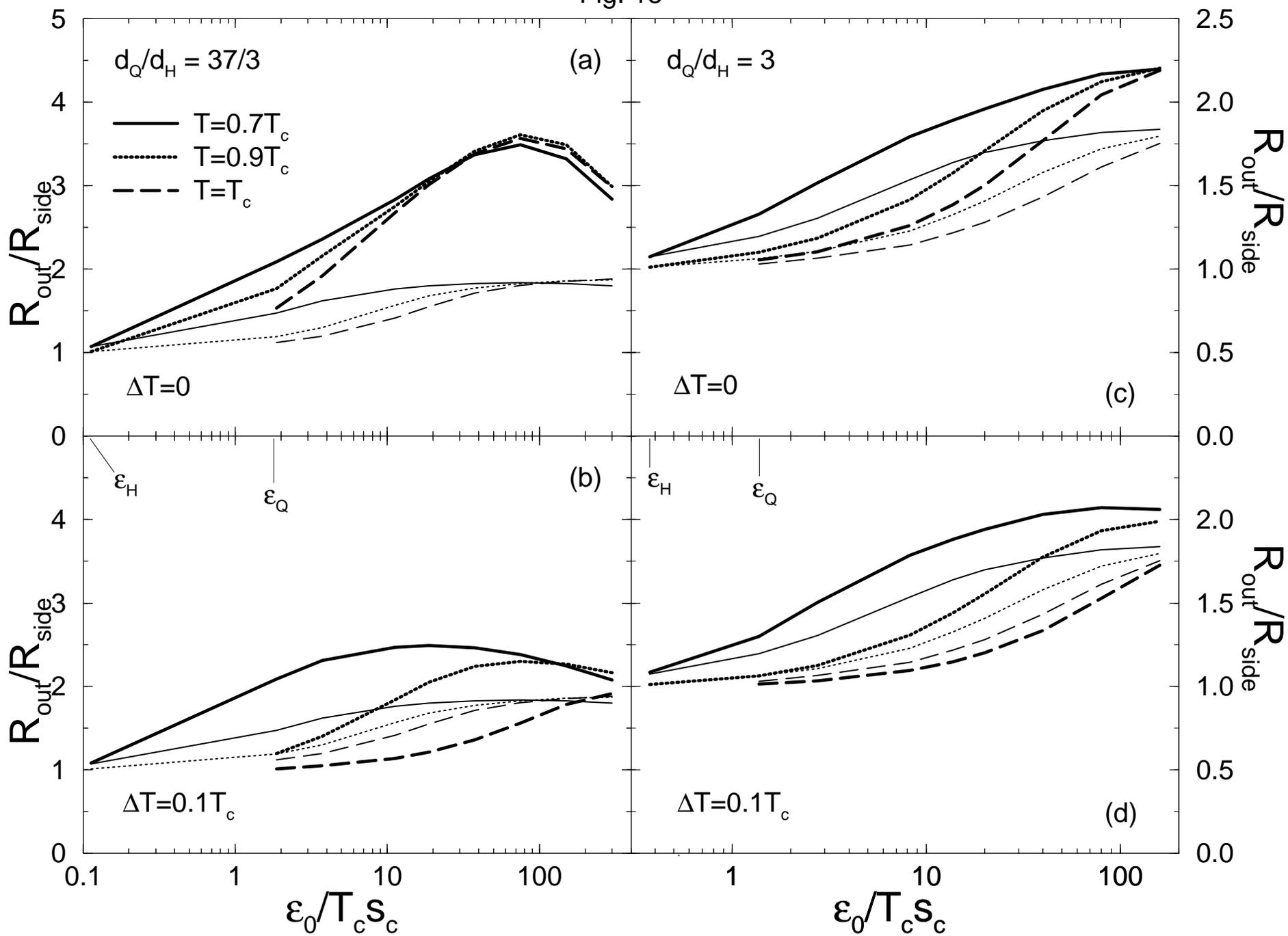
{\bf Fig.\ 18:} The same as in Fig.\ 13, but for the ratio
$R_{\rm \,out}/R_{\rm \,side}$.
\end{document}